\documentclass[fleqn,usenatbib]{mnras}

\usepackage{newtxtext,newtxmath}

\usepackage[T1]{fontenc}

\DeclareRobustCommand{\VAN}[3]{#2}
\let\VANthebibliography\thebibliography
\def\thebibliography{\DeclareRobustCommand{\VAN}[3]{##3}\VANthebibliography}

\usepackage{graphicx}	
\usepackage{amsmath}	
\usepackage{url}
\usepackage{orcidlink}

\title[D\&D 1\%]{DESI and DECaLS (D\&D): galaxy-galaxy lensing measurements with 1\% survey and its forecast}

\author[J. Yao et al.]{
Ji Yao$^{1,2,3}$\thanks{E-mail: ji.yao@outlook.com (JY)}\orcidlink{https://orcid.org/0000-0002-7336-2796},
Huanyuan Shan$^{1,4}$\thanks{E-mail: hyshan@shao.ac.cn (HS)}\orcidlink{https://orcid.org/0000-0001-8534-837X},
Pengjie Zhang$^{2,3,5}$\thanks{E-mail: zhangpj@sjtu.edu.cn (PZ)},
Eric Jullo$^{6}$\orcidlink{https://orcid.org/0000-0002-9253-053X},
Jean-Paul Kneib$^{6,7}$,
Yu Yu$^{2,3}$\orcidlink{https://orcid.org/0000-0002-9359-7170},
\newauthor
Ying Zu$^{2,3}$\orcidlink{https://orcid.org/0000-0001-6966-6925},
David Brooks$^{8}$,
Axel de la Macorra$^{9}$,
Peter Doel$^{8}$,
Andreu Font-Ribera$^{10}$\orcidlink{https://orcid.org/0000-0002-3033-7312},
\newauthor
Satya Gontcho A Gontcho$^{11}$\orcidlink{https://orcid.org/0000-0003-3142-233X},
Theodore Kisner$^{11}$\orcidlink{https://orcid.org/0000-0003-3510-7134},
Martin Landriau$^{11}$\orcidlink{https://orcid.org/0000-0003-1838-8528},
Aaron Meisner$^{12}$\orcidlink{https://orcid.org/0000-0002-1125-7384},
\newauthor
Ramon Miquel$^{13,10}$,
Jundan Nie$^{14}$\orcidlink{https://orcid.org/0000-0001-6590-8122},
Claire Poppett$^{11,15,16}$,
Francisco Prada$^{17}$\orcidlink{https://orcid.org/0000-0001-7145-8674},
Michael Schubnell$^{18,19}$,
\newauthor
Mariana Vargas Magana$^{9}$, and
Zhimin Zhou$^{14}$\orcidlink{https://orcid.org/0000-0002-4135-0977}
\\
$^{1}$Shanghai Astronomical Observatory (SHAO), Nandan Road 80, Shanghai, China\\
$^{2}$Department of Astronomy, School of Physics and Astronomy, Shanghai Jiao Tong University, Shanghai, China\\
$^{3}$Key Laboratory for Particle Astrophysics and Cosmology
(MOE)/Shanghai Key Laboratory for Particle Physics and Cosmology,
China\\
$^{4}$ University of Chinese Academy of Sciences, Beijing,  China\\
$^{5}$Tsung-Dao Lee Institute, Shanghai Jiao Tong University, Shanghai, China\\
$^{6}$Aix-Marseille Univ, CNRS, CNES, LAM, Marseille, France\\
$^{7}$Institute of Physics, Laboratory of Astrophysics, Ecole Polytechnique Fédérale de Lausanne (EPFL), Observatoire de Sauverny, 1290 Versoix, Switzerland\\
$^{8}$Department of Physics \& Astronomy, University College London, Gower Street, London, WC1E 6BT, UK\\
$^{9}$Instituto de F\'{\i}sica, Universidad Nacional Aut\'{o}noma de M\'{e}xico,  Cd. de M\'{e}xico  C.P. 04510,  M\'{e}xico\\
$^{10}$Institut de F\'{i}sica d’Altes Energies (IFAE), The Barcelona Institute of Science and Technology, Campus UAB, 08193 Bellaterra Barcelona, Spain\\
$^{11}$Lawrence Berkeley National Laboratory, 1 Cyclotron Road, Berkeley, CA 94720, USA\\
$^{12}$NSF's NOIRLab, 950 N. Cherry Ave., Tucson, AZ 85719, USA\\
$^{13}$Instituci\'{o} Catalana de Recerca i Estudis Avan\c{c}ats, Passeig de Llu\'{\i}s Companys, 23, 08010 Barcelona, Spain\\
$^{14}$National Astronomical Observatories, Chinese Academy of Sciences, A20 Datun Rd., Chaoyang District, Beijing, 100012, P.R. China\\
$^{15}$Space Sciences Laboratory, University of California, Berkeley, 7 Gauss Way, Berkeley, CA  94720, USA\\
$^{16}$University of California, Berkeley, 110 Sproul Hall \#5800 Berkeley, CA 94720, USA\\
$^{17}$Instituto de Astrof\'{i}sica de Andaluc\'{i}a (CSIC), Glorieta de la Astronom\'{i}a, s/n, E-18008 Granada, Spain\\
$^{18}$Department of Physics, University of Michigan, Ann Arbor, MI 48109, USA\\
$^{19}$University of Michigan, Ann Arbor, MI 48109, USA
}

\date{Accepted XXX. Received YYY; in original form ZZZ}

\pubyear{2015}

\begin{document}
\label{firstpage}
\pagerange{\pageref{firstpage}--\pageref{lastpage}}
\maketitle

\begin{abstract}
The shear measurement from DECaLS (Dark Energy Camera Legacy Survey) provides an excellent opportunity for galaxy-galaxy lensing study with DESI (Dark Energy Spectroscopic Instrument) galaxies, given the large ($\sim 9000$ deg$^2$) sky overlap. We explore this potential by combining the DESI  1\% survey and DECaLS DR8. With $\sim 106$ deg$^2$ sky overlap, we achieve significant detection of galaxy-galaxy lensing for BGS and LRG as lenses. {Scaled to the full BGS sample, we expect the statistical errors to improve from $18(12)\%$ to a promising level of $2(1.3)\%$ at $\theta>8^{'}(<8^{'})$. This brings stronger requirements for future systematics control.}
To fully realize such potential, we need to control the residual multiplicative shear bias $|m|<0.01$ and the bias in the mean redshift $|\Delta z|<0.015$. We also expect significant detection of galaxy-galaxy lensing with DESI LRG/ELG full samples as lenses, and cosmic magnification of ELG through cross-correlation with low-redshift DECaLS shear.  {If such systematical error control can be achieved,} we find the advantages of DECaLS, comparing with KiDS (Kilo Degree Survey) and HSC (Hyper-Suprime Cam),  are at low redshift, large-scale, and in measuring the shear-ratio (to $\sigma_R\sim 0.04$) and cosmic magnification. 
\end{abstract}




\section{Introduction}

Weak gravitational lensing is one of the most promising cosmological probes in studying the nature of dark matter, dark energy, and gravity \citep{Refregier2003,Mandelbaum2018}. The combination between different probes can be even more powerful, due to more constraining power and breaking the degeneracy between the parameters \citep{Planck2018I,DESY3cosmo}. However, possibly due to residual systematics or new physics beyond the standard $\Lambda$CDM model, the tension between CMB (cosmic microwave background) at redshift $z\sim1100$ and the late-time galaxy surveys at $z<\sim1$ {troubles us when using their synergy} \citep{Hildebrandt2017,HSC_Hamana2019,HSC_Hikage2019,Asgari2021,Heymans2021,DESY3cosmo,DESY3model,DESY3data,Planck2018I}. Many attempts have been made to 
examine this tension, in terms of different systematics \citep{Yamamoto2022,Wright2020,Yao2020,Yao2017,Kannawadi2019,Pujol2020,Mead2021,DESY3model,Amon2022,Fong2019}, different statistics \citep{Asgari2021,Joachimi2021,Lin2017b,Harnois-Deraps2021,Shan2018,Sanchez2021,Leauthaud2022,Chang2019}, and possible new physics \citep{Jedamzik2021}. We also refer to recent reviews for the readers' references \citep{Perivolaropoulos2021,Mandelbaum2018}.

To fully understand the physics behind this so-called ``$S_8$'' tension, different cosmological probes are required, as their sensitivities to the systematics are different. Many new observations are also needed, to explore different redshift ranges, sky patches, and even equipment properties. Among the many proposed stage IV galaxy surveys like Dark Energy Spectroscopic Instrument (DESI \cite{DESI2016a,DESI2016b}), Vera C. Rubin Observatory’s Legacy Survey of Space and Time (LSST, \citealt{LSST2009}), Euclid \citep{Euclid2011}, Roman Space Telescope (or WFIRST, \citealt{WFIRST2015}) and China Space Station Telescope (CSST, \citealt{Gong2019}), DESI is the only one currently operating {and 
has measured more than 7.5 million redshifts so far}.

DESI itself will provide tremendous constraining power in studying the expansion history of the Universe as well as the large-scale structure \citep{DESI2016a}. Its cross-correlations with other lensing surveys (referred to as galaxy-galaxy lensing or g-g lensing) will provide not only more, but also independent cosmological information \citep{Prat2021,Joudaki2018,Sanchez2021}, while it can be used to study the galaxy-matter relation \citep{Leauthaud2022,Leauthaud2017}, test gravity \citep{Zhang2007,Jullo2019,Blake2020}, and study the systematics \citep{Yao2020,Yao2017,SC2008,Zhang2010photoz,Giblin2021}. However, stage III surveys like DES (Dark Energy Survey, \citealt{DESY3cosmo}), KiDS (Kilo-Degree Survey, \citealt{Heymans2021}), and HSC (Hyper-Suprime Cam, \citealt{HSC_Hikage2019}) do not offer extremely large overlap with DESI, while the stage IV surveys mentioned previously will require many years of observations before reaching their full overlap with DESI. In short, the sky overlap will limit the cross-correlation studies with DESI in the near future.

In this work, we study the cross-correlations between galaxy shear measured from DECaLS (Dark Energy Camera Legacy Survey) DR8 and galaxies from the DESI $1\%$ (SV3) survey, and compare those with the overlapped data from KiDS and HSC. We measure the g-g lensing signals of the different weak lensing surveys with DESI 1\% survey and estimate their S/N (signal-to-noise ratio) that can be achieved with full DESI in the future. We explore the advantages of DECaLS, and exhibit the measurements of shear-ratio and cosmic magnification as two promising tools in using the great constraining power of DECaLS $\times$ DESI. Additionally, to achieve the expected precision, we propose requirements on the DECaLS data, in terms of the shear calibration and the redshift distribution calibration.

This work is organized as follows. In Section\,\ref{sec theory} we briefly introduce the observables and their theoretical predictions.  In Section\,\ref{sec data} we describe the DESI, DECaLS, KiDS, and HSC data we use. In Section\,\ref{sec results} we show the g-g lensing measurements for different DESI density tracers and different lensing surveys, and the measurements of shear-ratio and cosmic magnification. We summarize our findings from DESI$\times$DECaLS for the 1\% survey in Section.\,\ref{sec conclusions}.



\section{Theory} \label{sec theory}

In this section, we briefly review the theory of the g-g lensing observables. We assume spacial curvature $\Omega_k=0$ so that the comoving radial distance equals the comoving angular diameter distance.

\subsection{Galaxy-galaxy lensing}
\label{sec g-g lensing}

Since the foreground gravitational field can distort the shape of the background galaxy, there will be a correlation between the background galaxies' gravitational shear $\gamma^{\rm G}$ and the foreground galaxies' number density $\delta_{\rm g}$. {The correlation of $\left<\delta_{\rm g}\gamma^{\rm G}\right>$ (or $w^{\rm gG}$) will probe the clustering of the underlying matter field $\left<\delta_{\rm m}\delta_{\rm m}\right>$ (or the matter power spectrum $P_{\rm \delta}(k)$), the galaxy bias $b_g(k,z)$, and the redshift-distance relation, which are sensitive to the cosmological model and gravitational theory.} We recall the g-g lensing angular power spectrum \citep{Prat2021}:
\begin{equation}
C^{g\kappa}(\ell)=\int_{0}^{\chi_{\rm max}}\frac{n_{\rm l}(\chi)q_{\rm s}(\chi)}{\chi^2} b_{\rm g}(k,z) P_{\rm \delta}\left(k=\frac{\ell+1/2}{\chi},z\right)d\chi, \label{eq C^gG}
\end{equation}
which is a weighted projection from the 3D {non-linear} matter power spectrum $P_{\rm \delta}(k,z)$ to the 2D galaxy-lensing convergence angular power spectrum $C^{g\kappa}(\ell)$. It will also depend on the galaxy bias $b_{\rm g}=\delta_{\rm g}/\delta_{\rm m}$, the comoving distance $\chi$, the redshift distribution of the lens galaxies $n_{\rm l}(\chi)=n_{\rm l}(z)dz/d\chi$, and the lensing efficiency as a function of the lens position (given the distribution of the source galaxies) $q_{\rm s}(\chi)$, which is written as
\begin{equation}
q_{\rm s}(\chi_{\rm l}) = \frac{3}{2}\Omega_{\rm m}\frac{H_0^2}{c^2}(1+z_{\rm l})
	\int_{\chi_{\rm l}}^\infty
	n_{\rm s}(\chi_{\rm s})\frac{(\chi_{\rm s}-\chi_{\rm l})\chi_{\rm l}}{\chi_{\rm s}}d\chi_{\rm s}, \label{eq q}
\end{equation}
where $n_{\rm s}(\chi_{
\rm s})$ denotes the distribution of the source galaxies as a function of comoving distance, while $\chi_{\rm s}$ and $\chi_{\rm l}$ denote the comoving distance to the source and the lens, respectively.

The real-space galaxy-shear correlation function can be obtained through the Hankel transformation
\begin{equation}
w^{\rm gG}(\theta) = \frac{1}{2\pi}\int_{0}^{\infty}d\ell \ell C^{g\kappa}(\ell) J_2(\ell\theta) \label{eq w Hankel},
\end{equation}
where $J_2(x)$ is the Bessel function of the first kind with order 2. The ``G'' represents the gravitational lensing shear $\gamma^{\rm G}$, which is conventionally used to separate from the intrinsic alignment $\gamma^{\rm I}$, whose contribution is ignored in this work due to the photo-$z$ separation shown later. 

Therefore, by observing the correlation of $w^{\rm gG}$, we can derive the constraints on the cosmological parameters through Eq.\,\eqref{eq C^gG}, $P_{\rm \delta}(k)$ and $\chi(z)$. In order to get the precise cosmology, many systematics need to be considered, for example, the shear calibration error that can shift the measurement of $w^{\rm gG}$, the inaccurate estimation of redshift distribution for the source $n_{\rm s}(\chi_{\rm s}(z_{\rm s}))$ which can bias the theoretical estimation of Eq.\,\eqref{eq C^gG}, the massive neutrino effects and the baryonic effects that can bias the matter power spectrum $P_{\rm \delta}(k,z)$, {and the non-linear galaxy bias $b_g(k,z)$}\footnote{In this work we use the mathematical classification of linear/non-linear bias as a matched filter, however, for more physical modeling, this is normally expressed as 1-halo/2-halo terms and HOD (halo occupation distribution) descriptions such as central/satellite fractions \citep{Leauthaud2017}}. In this work, we mainly focus on the statistical significance for DESI$\times$DECaLS, rather than the systematics. The current statistical error for the $1\%$ survey is expected to be more dominant, but for cautious reasons, we will not give final estimations on the cosmological parameters. 

\subsection{Shear-ratio} \label{sec shear-ratio}

The g-g lensing two-point statistics normally contain stronger detection significance at the small-scale than at the large-scale, due to a stronger tidal gravitational field and more galaxy pairs (throughout the whole sky, not around a particular galaxy). However, due to the inaccurate modeling of small-scale effects, such as the non-linear galaxy bias $b_{\rm g}(k,z)$, suppression in the matter power spectrum $P_{\rm \delta}(k)$ due to massive neutrino and baryonic effects, etc., the small-scale information is conventionally abandoned \citep{Heymans2021,DESY3cosmo,Lee2022}. However, by choosing the same lens galaxies with source galaxies at different redshifts, i.e. with the same redshift distribution $n_u(z)$ for the lens while different redshift distribution $n_v(z)$ and $n_w(z)$ for the sources, the ratio between the angular power spectra $C^{g\kappa}_{uv}$ and $C^{g\kappa}_{uw}$ (or the correlation functions $w^{\rm gG}_{uv}$ and $w^{\rm gG}_{uw}$) will mainly base on the two lensing efficiency functions as in Eq.\,\eqref{eq q} for the $v$-th and $w$-th source bins. This ratio does not suffer strongly from the modeling of the galaxy bias $b_{\rm g}$ or the matter power spectrum $P_{\rm \delta}(k)$, as they share the same lens sample according to Eq.\,\eqref{eq C^gG}. {The shear-ratio (or lensing-ratio) has been used to improve cosmological constraints \citep{Sanchez2021}, as it is sensitive to the $\chi(z)$ relation in Eq.\,\eqref{eq q} and the nuisance parameters for the systematics,} or to study the shear bias \citep{Giblin2021}. In this work, we will show the great potential of measuring shear-ratio with DESI$\times$DECaLS.

To account for the full covariance in measuring shear-ratio $R=w_2/w_1$, and to prevent possible singular values when taking the ratio (when $w_1\sim0$), we construct the following data vector \begin{equation}
    V=w_1R-w_2, \label{eq constructed vector}
\end{equation}
which is designed to be $0$ when $R$ is correctly predicted from the two data sets $w_1$ and $w_2$ that we want to take the ratio. The resulting covariance for the data vector $V$ is
\begin{equation}
    C' = R^2 C_{11} + C_{22} - R(C_{12}+C_{21}), \label{eq constructed cov}
\end{equation}
where $C_{ij}$ is the covariance between $w_i$ and $w_j$. The likelihood of $-2{\rm ln}\mathscr{L}=V^{\rm T}C'^{-1}V$ will give the posterior of the shear-ratio $R$. To account for the covariance is $R$-dependent, normalization is done thereafter {so that its PDF satisfies $\int P(R)dR=1$}. An alternative way is to marginalize over the theoretical predictions $w_i$, similar to \cite{Sun2022,Dong2022}, which we leave for future studies.

\subsection{Cosmic magnification}
\label{sec mag theory}

The observed galaxy number density is affected by its foreground lensing signals, leading to an extra fluctuation besides the intrinsic clustering of galaxies, namely,
\begin{equation}
\delta_{\rm g}^{\rm L} = \delta_{\rm g} + g_{\mu}\kappa, \label{eq magnification}
\end{equation}
where $\delta_{\rm g}^{\rm L}$ denotes the observed lensed galaxy overdensity, $\delta_{\rm g}$ denotes the intrinsic overdensity of galaxies due to gravitational clustering, $\kappa$ is the lensing convergence affecting the flux and the positions of the foreground galaxy sample, and due to the foreground inhomogeneities. For a {complete and} flux-limited sample, the magnification amplitude $g_\mu=2(\alpha-1)$. In that case, the magnification amplitude is sensitive to the galaxy flux function $N(F)$, denoting the number of galaxies brighter than flux limit $F$, with $\alpha=-d{\rm ln}N/d{\rm ln}F$.

According to Eq.\,\eqref{eq magnification}, for a given galaxy sample at $z=z_1$, it not only contains clustering information of $\delta_{\rm g}(z=z_1)$, but also has lensing information of $\kappa$ from the matter at $z<z_1$, which is normally treated as a contamination to the clustering signals \citep{vonWietersheim-Kramsta2021,Deshpande2020,Kitanidis2021}. Meanwhile, attempts have been made to directly measure the cosmic magnification as a source of cosmological information \citep{Liu2021,Gonzalez-Nuevo2020,Yang2017}.

We follow the method of \cite{Liu2021} and correlate the shear galaxies at lower redshift (bin $i$) and the number density galaxies at higher redshift (bin $j$),
\begin{equation}
C^{\kappa\mu}_{ij}(\ell)=g_\mu \int_{0}^{\chi_{\rm max}}\frac{q_i(\chi)q_j(\chi)}{\chi^2}  P_{\rm \delta}\left(k=\frac{\ell+1/2}{\chi},z\right)d\chi, \label{eq C^muG}
\end{equation}
which requires the redshift distribution of $n_i(z)$ being significantly separated from $n_j(z)$, so that the intrinsic clustering $\times$ lensing shear signal vanishes. The corresponding correlation function from the Hankel transformation is similar to Eq.\,\eqref{eq w Hankel}.

\subsection{Signal-to-noise definition}
\label{sec S/N}

The S/N definition in this work uses amplitude fitting. For a given measurement $w_{\rm data}$ and an assumed theoretical model $w_{\rm model}$, we fit an amplitude $A$ to the likelihood:
\begin{equation}
    -2{\rm ln}\mathscr{L}=\left(w_{\rm data}-Aw_{\rm model}\right){\rm Cov}^{-1}\left(w_{\rm data}-Aw_{\rm model}\right),
\end{equation}
so that a posterior of $A^{+\sigma_A}_{-\sigma_A}$ can be obtained, {where $\sigma_A$ is the Gaussian standard deviation}. Then the corresponding S/N is $A/\sigma_A$.

We note that, if $w_{\rm data}$ is a single value rather than a data vector, this S/N defined by amplitude fitting is identical to the S/N of the data itself, namely $A/\sigma_A=w_{\rm data}/\sigma_{w_{\rm data}}$. This is the case for most of the S/N calculated in this work, when there is one single measurement at small-scale and one at large-scale, and the small-scale and large-scale data correspond to different (nonlinear/linear) galaxy biases so they should be treated separately.

\section{Data} \label{sec data}
In this section, we introduce the DESI spectroscopic data and the shear catalogs from DECaLS/KiDS/HSC. We note even though the DES-Y3 catalog can have an overlap with full DESI for $\sim1264$ deg$^2$, its overlap with DESI SV3 catalog is $0$. We, therefore, do not present any analysis for DES.

\subsection{DESI} \label{sec DESI}
DESI is the only operating Stage IV galaxy survey. It is designed to cover 14,000 deg$^2$ of the sky, with 5,000 fibers collecting spectra simultaneously \citep{DESI2016b,Silber2022,Miller2022}. DESI aims to observe density tracers such as BGS (Bright Galaxy Survey, \citealt{RuizMacias2020}), LRG (luminous red galaxies, \citealt{Zhou2020}), ELG (emission line galaxies, \citealt{Raichoor2020}), and QSO (quasi-stellar objects, \citealt{Yeche2020}), with generally increasing redshift. Other supporting papers on target selections and validations can be find in \cite{AllendePrieto2020,Alexander2022,Lan2022,Cooper2022,Hahn2022,Zhou2022,Chaussidon2022}. DESI plans to use these tracers to study cosmology, especially in BAO (baryonic acoustic oscillations) and RSD (redshift-space distortions) \citep{DESI2016a,Levi2013}. It is located on the 4-meter Mayall telescope in Kitt Peak, Arizona \citep{DESI2022a}. From 2021 till now, DESI has finished its ``SV3'' \citep{DESIsv} and ``DA0.2'' catalogs, which will be included in the coming Early Data Release (EDR, \citealt{DESIdr}). The Siena Galaxy Atlas \citep{DESIsga} is also expected soon.

The DESI experiment is based on the DESI Legacy Imaing Surveys \citep{Zou2017,Dey2019,Schlegel2022}, with multiple supporting pipelines in spectroscopic reduction \citep{Guy2022}, derivation of classifications and redshifts \citep{Bailey2022}, fiber assigement \citep{Raichoor2022}, survey optimization \citep{Schlafly2022}, spectroscopic target selection \citep{Myers2022}

In this work, we use the DESI SV3 catalog, which is also known as the $1\%$ survey (with a sky coverage of $\sim140$ deg$^2$), for the g-g lensing study. We consider the DESI BGS, LRGs, and ELGs, while ignoring the QSOs as the available number is relatively low. {In SV3, each galaxy is assigned a weight to account for the survey completeness and redshift failure.} Since the purpose of this paper is not a precise measurement of cosmology, we assume the linear galaxy biases follow $b_{\rm BGS}(z)D(z)=1.34$, $b_{\rm LRG}(z)D(z)=1.7$, and $b_{\rm ELG}(z)D(z)=0.84$, where $D(z)$ is the linear growth factor normalized to $D(z=0)=1$ \citep{DESI2016a}. The number of galaxies used will be informed later in the paper, as the overlap between the DESI $1\%$ survey and the lensing surveys are different.

\subsection{DECaLS} \label{sec DECaLS}
We use lensing shear measurement from DECaLS DR8, which contains galaxy images in $g-$, $r-$, and $z-$bands \citep{Dey2019}. DECaLS DR8 galaxies are processed by Tractor \citep{Meisner2017,Lang2014} and divided into five types according to their morphologies: PSF, SIMP, DEV, EXP, and COMP \citep{Phriksee2020,Yao2020,Zu2021,Xu2021}. The galaxy ellipticities $e_{1,2}$ are measured ---- except for the PSF type ---- with a joint fit on the $g-$, $r-$, and $z-$bands. A conventional shear calibration \citep{Heymans2012,Miller2013,Hildebrandt2017} is applied as in
\begin{equation} \label{eq shear calib}
\gamma^{\rm obs} = (1+m)\gamma^{\rm true}+c,
\end{equation}
with a multiplicative bias $m$ and additive bias $c$, to account for possible residual bias from PSF modeling, measurement method, blending and crowding \citep{Mandelbaum2015,Martinet2019}. This calibration is obtained by comparing with Canada–France–Hawaii Telescope (CFHT) Stripe 82 observed galaxies and Obiwan simulated galaxies \citep{Phriksee2020,Kong2020}.

Several versions of the photometric redshift for the DECaLS galaxies have been estimated \citep{Zou2019,Zhou2021,Duncan2022}. We apply the most widely used one \citep{Zhou2021}, which uses the $g$, $r$, and $z$ optical bands from DECaLS while borrowing $W1$ and $W2$ infrared bands from WISE (Wide-field Infrared Survey Explorer, \citealt{Wright2010}). The photo-$z$ algorithm is trained based on a decision tree, with training samples constructed from a wide selection of spectroscopic redshift surveys and deep photo-$z$ surveys. We additionally require $z<21$ to select galaxies with better photo-$z$. {We use the photo-z distribution to represent the true-z distribution $n(z)$, while allowing a systematic bias of $\Delta z$ in the form $n(z-\Delta z)$, to pass its effect to Eq.\,\eqref{eq q} then Eq.\,\eqref{eq C^gG}. This is appropriate as weak lensing is mainly biased due to the mean redshift but slightly affected by the redshift scatter.}

Overall, the DR8 shear catalog has $\sim 9,000$ deg$^2$ sky coverage ---- which will be the final overlap with full DESI ---- with an average galaxy number density of $\sim1.9$ gal/arcmin$^2$. The overlapped area with DESI $1\%$ survey is $\sim106$ deg$^2$, which is significantly larger than the other stage III lensing surveys.

We note that the current DECaLS DR8 shear catalog can have some residual multiplicative bias $|m|\sim0.05$ \citep{Yao2020,Phriksee2020}, {possibly due to the selections in observational data while making the comparison \citep{Li2020,Jarvis2016}.} This will prevent us from getting reliable cosmology for measurements with $S/N>\sim20$. Also, there exists a possible bias in the redshift distribution $n(z)$, which will require a galaxy color-based algorithm \citep{Hildebrandt2017,Buchs2019,Wright2020} or a galaxy clustering-based algorithm \citep{Peng2022,Zhang2010photoz,vandenBusch2020} to get the correction. For these two reasons, we choose not to extend this study to the precision cosmology level. A future version of the DECaLS DR9 shear catalog is under development, with improved data reduction and survey procedures\footnote{https://www.legacysurvey.org/dr9/description/}, with more advanced shear calibration {for a pure Obiwan image simulation-based algorithm} (Yao et al. in preparation) and redshift calibration (Xu et al. in preparation).

\subsection{KiDS}
The Kilo-Degree Survey is run by the European Southern Observatory and is designed for weak lensing studies in $ugri$ optical bands. The KiDS data are processed by THELI \citep{Erben2013} and Astro-WISE \citep{deJong2015,Begeman2013}. The galaxy shear measurements are obtained by $lens$fit \citep{Conti2017,Miller2013}, and the photo-$z$s are measured by BPZ \citep{Benitez2000,Benitez2004} using the KiDS $ugri$ optical bands and the $ZYJHK_{\rm s}$ infrared bands from VIKING \citep{Wright2019}. The KiDS shears are calibrated following the same equation as Eq.\,\eqref{eq shear calib} with image simulation \cite{Kannawadi2019}.

We use the KiDS-1000 shear catalog \citep{Giblin2021,Asgari2021} in this work. The overlapped area with DESI SV3 is $\sim55$ deg$^2$. The expected overlapped area between the full DESI footprint and KiDS-1000 is $\sim456$ deg$^2$.

\subsection{HSC}
The Hyper Suprime-Cam Subaru Strategic Program (HSC-SSP, or HSC) is a Japanese lensing survey using the powerful Subaru telescope. It covers five photometric bands $grizy$. Compared with KiDS and DES, HSC has its unique advantage in the galaxy number density and high-z galaxies (but with a smaller footprint). The HSC shears are calibrated similarly to Eq.\,\eqref{eq shear calib} \citep{Mandelbaum2018HSC} but with an additional shear responsivity \citep{HSC_Hamana2019}.

We use the HSC-Y1 shear catalog \citep{HSC_Hikage2019,HSC_Hamana2019}, which overlaps with DESI SV3 for $\sim48$ deg$^2$. The expected overlap between HSC-Y3 data and full DESI is $\sim733$ deg$^2$.

\begin{figure}
	\includegraphics[width=\columnwidth]{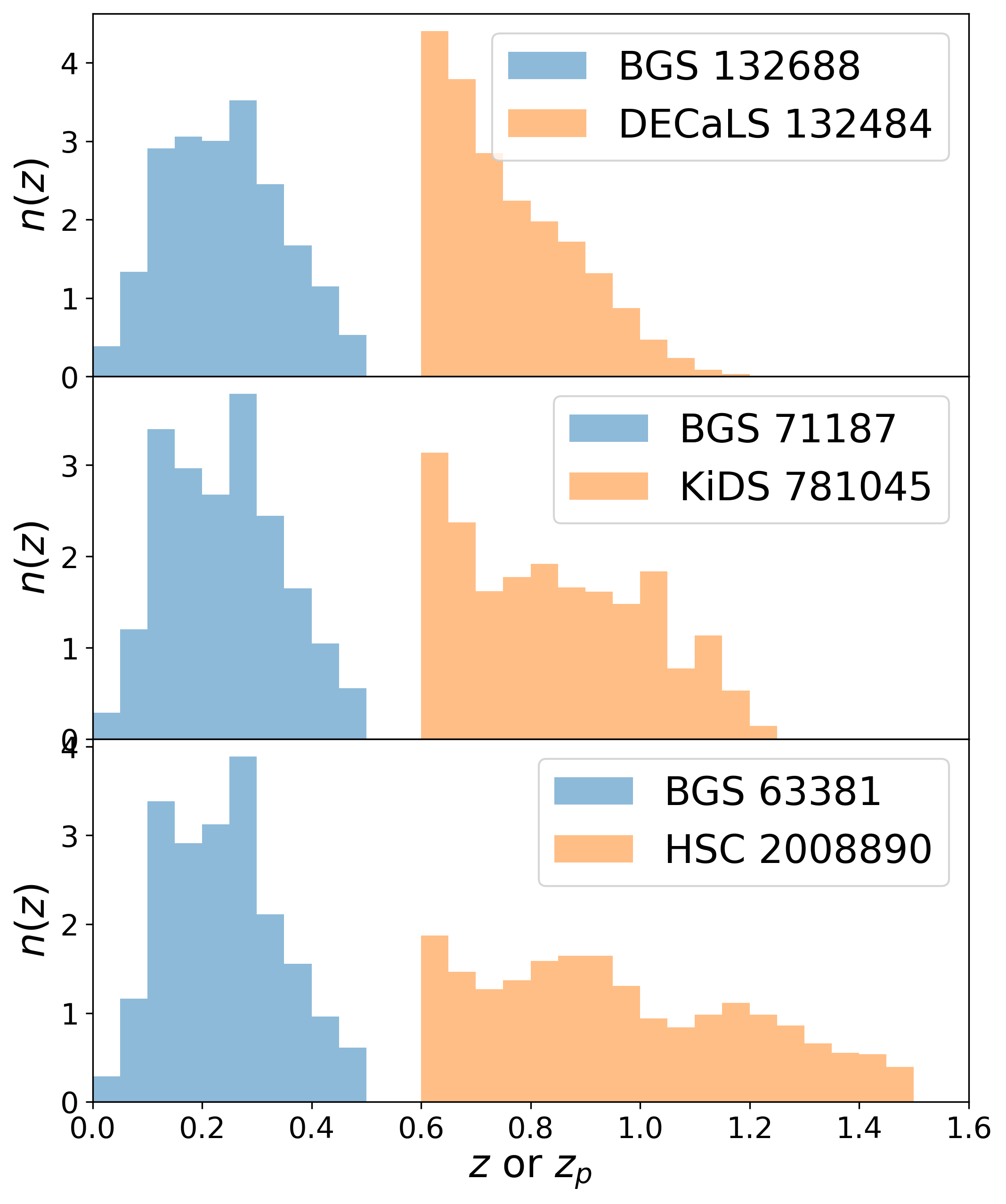}
    \caption{The galaxy redshift distributions for the DESI BGS with $0<z<0.5$ and photo-$z$ distributions for the lensing surveys with $0.6<z_p<1.5$. The numbers in the labels are the number of galaxies in the overlapped region.}
    \label{fig: nz BGS}
\end{figure}

\begin{figure}
	\includegraphics[width=\columnwidth]{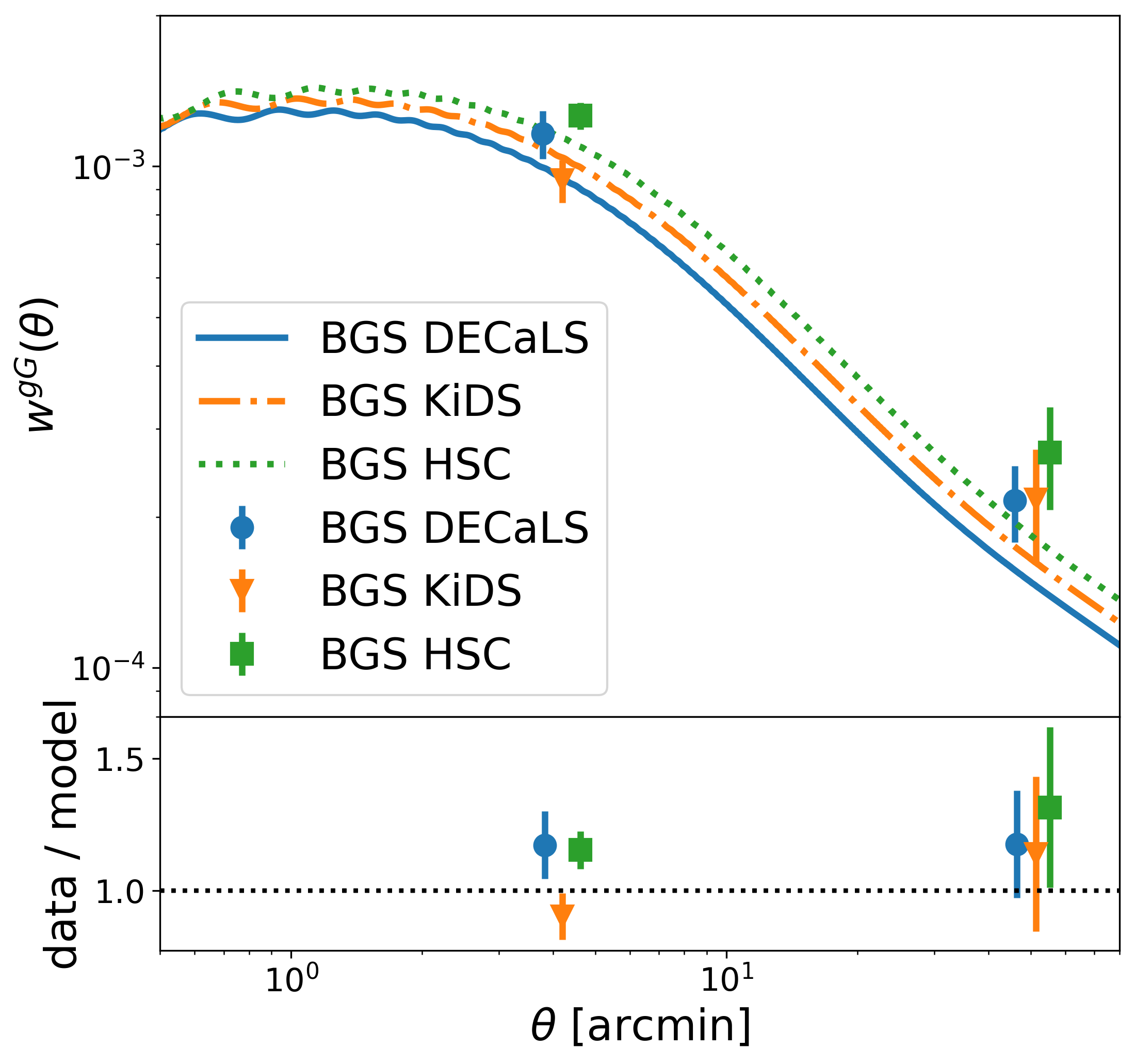}
    \caption{The galaxy-galaxy lensing angular correlation functions, corresponding to the galaxies samples in Fig.\,\ref{fig: nz BGS}. In the upper panel, the theoretical curves are given by the fiducial cosmology and the assumed galaxy bias model. The \{small-scale, large-scale\} detection significances are \{9.1, 5.8\} for BGS$\times$DECaLS, \{10.2, 3.9\} for BGS$\times$KiDS , and \{16.1, 4.3\} for BGS$\times$HSC. In the lower panel, we show the ratio between our measurements and the corresponding theoretical model, with the latter re-weighted using the number of pairs and lensing weights to account for the band power problem with wide angular bins. The DECaLS and HSC results are slightly shifted horizontally.}
    \label{fig: wgG BGS}
\end{figure}

\begin{figure}
	\includegraphics[width=\columnwidth]{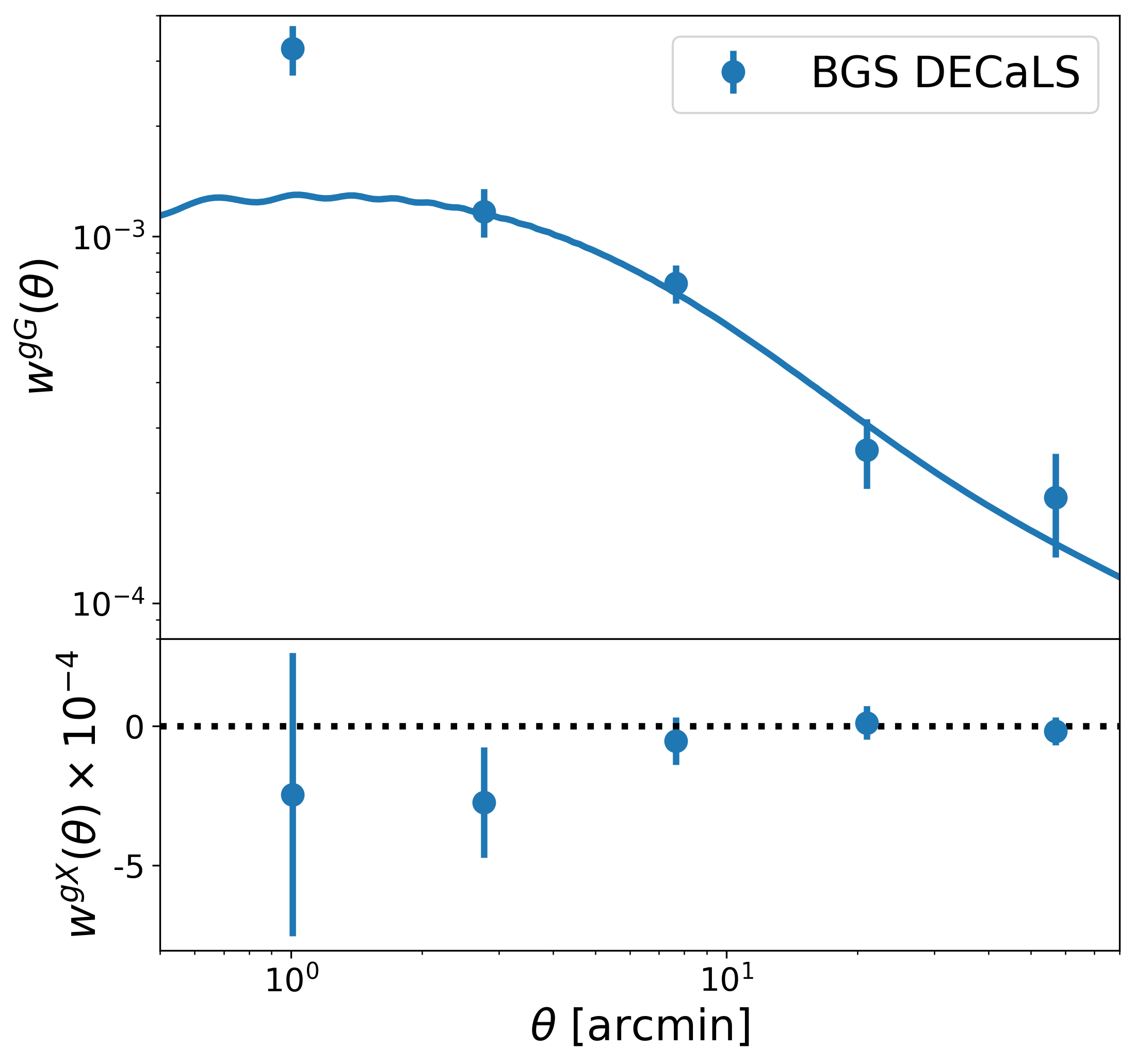}
    \caption{The galaxy-galaxy lensing angular correlation function $w^{\rm gG}$ (upper panel) and its $45\deg$-rotation test $w^{\rm gX}$ (lower panel) for the BGS$\times$DECaLS g-g lensing only, with the same distribution as in Fig.\,\ref{fig: nz BGS} but with more angular bins with 50 jackknife sub-regions. In the upper panel, the theoretical curves are given by the fiducial cosmology and the assumed galaxy bias model. The detection significance for the 5 angular bins are \{6.5, 6.6, 8.4, 4.7, 3.2\}, with the 4 large-scale bins well-agreed with the prediction from fiducial cosmology and the linear bias assumption. {The total S/N using amplitude fitting (as described in Sec.\,\ref{sec S/N}) is $8.9\sigma$ ($A=1.03^{+0.12}_{-0.11}$) for the right three large-scale dots, and is $10.0\sigma$ ($A=1.0^{+0.1}_{-0.1}$) for the right four large-scale dots.} In the lower panel where the shear are rotated for $45\deg$, the results are consistent with 0, with reduced-$\chi^2\sim3/5$.}
    \label{fig: wgG BGS DECaLS}
\end{figure}

\section{Results} \label{sec results}
In this section, we show the measurements of different galaxy-shear correlation functions. The estimator for the galaxy-shear correlation is: 
\begin{equation} \label{eq gG estimator}
w^{\rm gG }(\theta)=\frac{\sum_{\rm ED}\textsc{w}_{\rm E}\gamma^+_{\rm E}\textsc{w}_{\rm D}}{\sum_{\rm ER}(1+m_{\rm E})\textsc{w}_{\rm E}\textsc{w}_{\rm R}}
-\frac{\sum_{\rm ER}\textsc{w}_{\rm E}\gamma^+_{\rm E}\textsc{w}_{\rm R}}{\sum_{\rm ER}(1+m_{\rm E})\textsc{w}_{\rm E}\textsc{w}_{\rm R}}\ ,
\end{equation}
where $\textsc{w}_{\rm E}$, $m_{\rm E}$ and $\gamma^+_{\rm E}$ denotes the lensing weight (inverse-variance weight for DECaLS \citealt{Phriksee2020} and HSC \citealt{HSC_Hikage2019}, an adjusted version for KiDS \citealt{Miller2013}), the multiplicative bias correction (for HSC there is an extra shear responsivity included), and the tangential shear of the source galaxy, with respect to the given lens galaxy with weight $\textsc{w}_{\rm D}$ or $\textsc{w}_{\rm R}$. The $\Sigma$-summations are calculated for all the ellipticity-density (ED) pairs and the ellipticity-random (ER) pairs. We note Eq.\,\eqref{eq gG estimator} already includes the correction for boost factor \citep{Mandelbaum2005boostfactor,Amon2018}, {and this equation is adequate for the multiplicative bias $m_{\rm E}$ defined either per galaxy or per sample.} The correlation uses DESI official random catalogs to simultaneously correct for the additive bias in the presence of a mask and reduce the shape noise. We will show the measurements with different lens samples and source catalogs using the above estimator.

\subsection{DESI $w^{\rm gG}$}
\label{sec wgG}

We first show the g-g lensing measurements for DESI BGS and the three shear catalogs. The normalized redshift distributions $n(z)$ are shown in Fig.\,\ref{fig: nz BGS}, with the number of galaxies being used in the labels. We use BGS with $0<z<0.5$, and require the photo-$z$ of the source galaxies located at $0.6<z_p<1.5$, so that the overlap in redshift is very small even considering the inaccuracy of photo-$z$. We see that DECaLS has the most available BGS lenses, while HSC has the most available sources and the highest redshift. We notice there are unexpected spikes for the photo-z distribution of KiDS, which is probably due to cosmic variance as the overlapped area is much smaller than the full KiDS data.

\begin{figure}
	\includegraphics[width=\columnwidth]{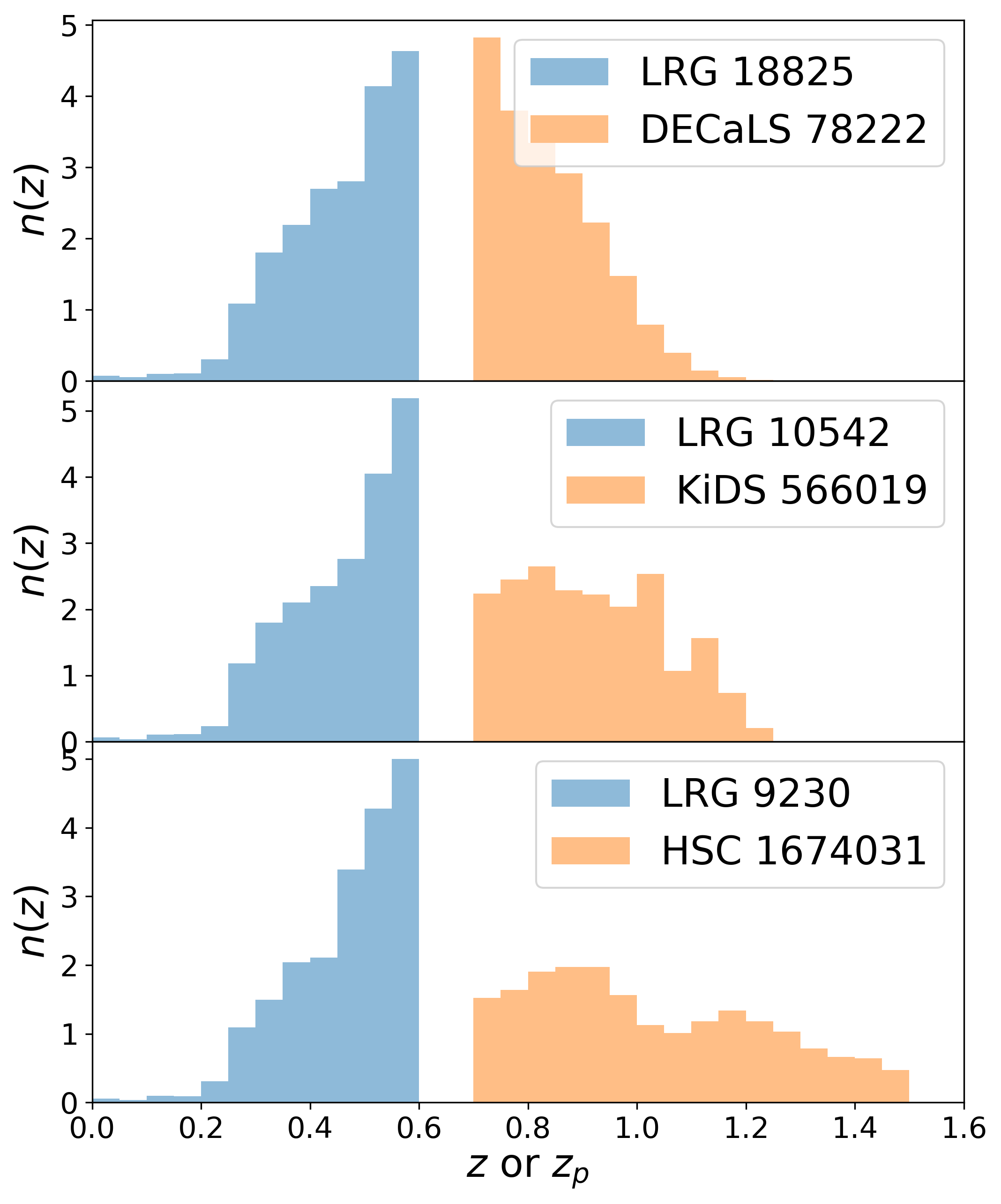}
    \caption{The galaxy redshift distributions for the DESI LRGs with $0<z<0.6$ and photo-$z$ distributions for the lensing surveys with $0.7<z_p<1.5$. The numbers in the labels are the number of galaxies in the overlapped region.}
    \label{fig: nz LRG}
\end{figure}

\begin{figure}
	\includegraphics[width=\columnwidth]{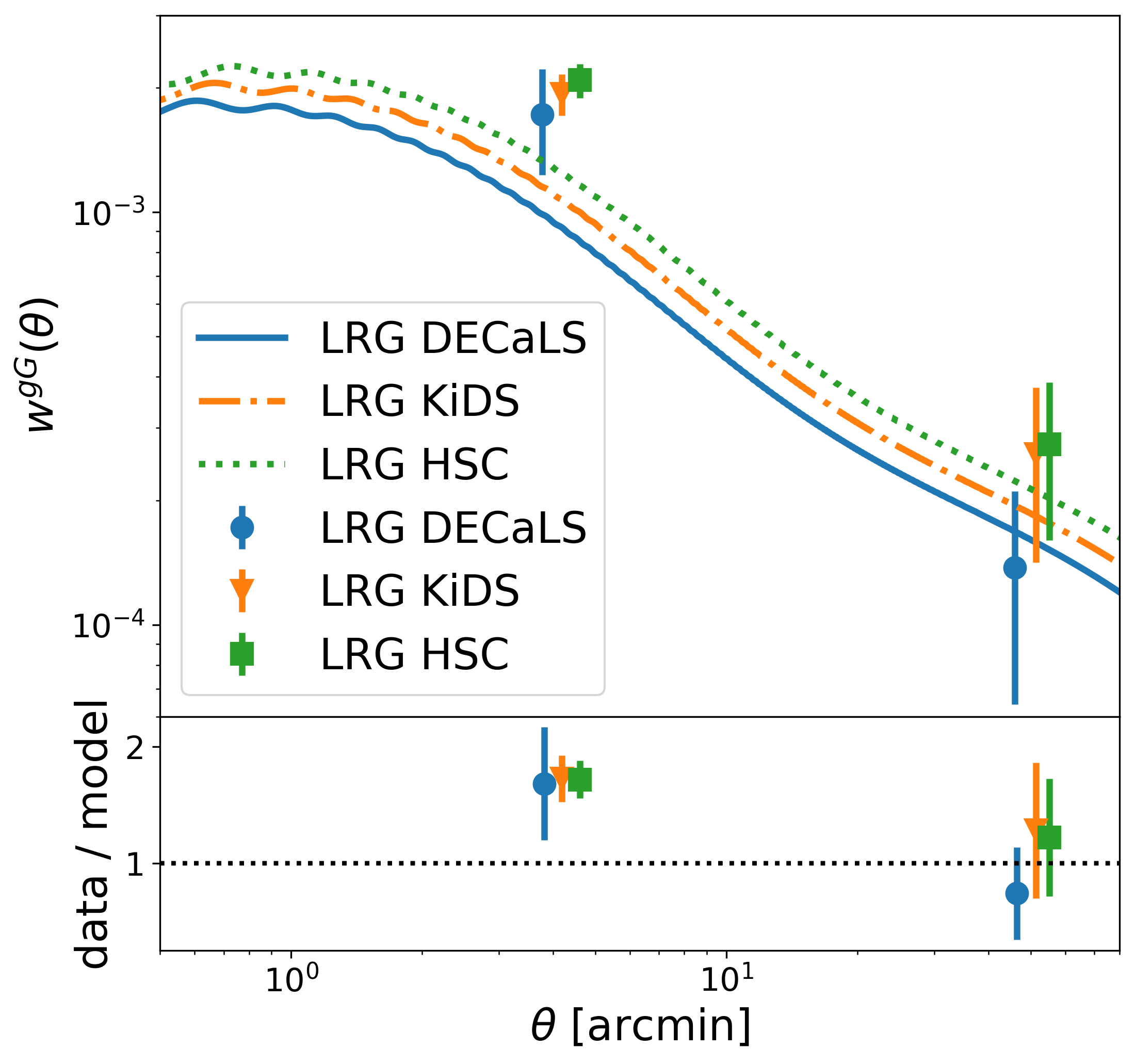}
    \caption{The galaxy-galaxy lensing angular correlation functions, corresponding to the galaxies samples in Fig.\,\ref{fig: nz LRG}. In the upper panel, the theoretical curves are given by the fiducial cosmology and the assumed galaxy bias model. The \{small-scale, large-scale\} detection significances are \{3.5, 1.9\} for LRG$\times$DECaLS,  \{8.7, 2.2\} for LRG$\times$KiDS, and \{10.6, 2.4\} for LRG$\times$HSC. }
    \label{fig: wgG LRG}
\end{figure}

\begin{figure}
	\includegraphics[width=\columnwidth]{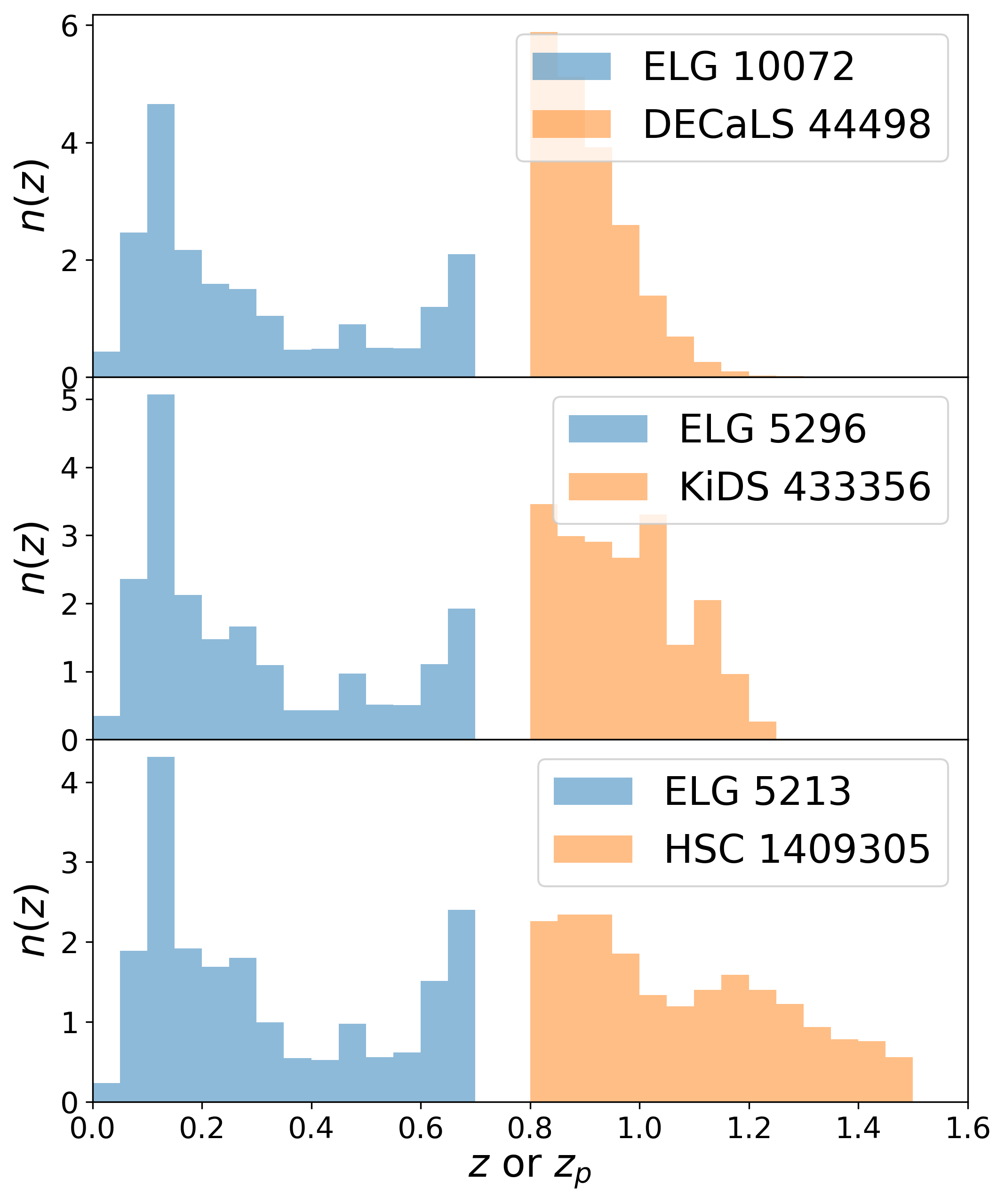}
    \caption{The galaxy redshift distributions for the DESI ELGs with $0<z<0.7$ and photo-$z$ distributions for the lensing surveys with $0.8<z_p<1.5$. The numbers in the labels are the number of galaxies in the overlapped region.}
    \label{fig: nz ELG}
\end{figure}

\begin{figure}
	\includegraphics[width=\columnwidth]{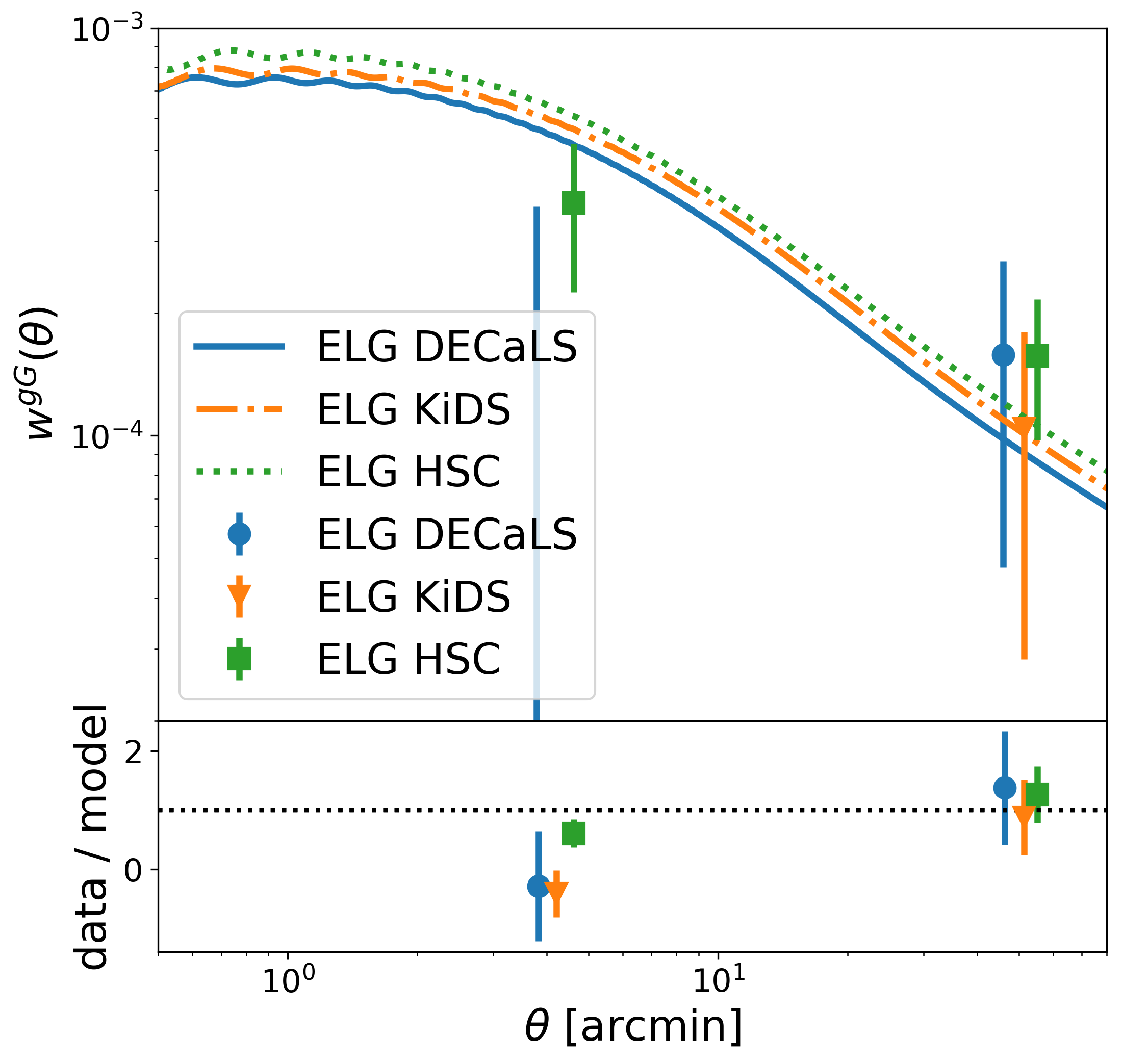}
    \caption{The galaxy-galaxy lensing angular correlation functions, corresponding to the galaxies samples in Fig.\,\ref{fig: nz ELG}. The theoretical curves are given by the fiducial cosmology and the assumed galaxy bias model. The \{small-scale, large-scale\} detection significance are \{-0.3, 1.4\} for ELG$\times$DECaLS, \{-1.1, 1.4\} for ELG$\times$KiDS, and \{2.5, 2.6\} for ELG$\times$HSC. The negative values at small-scale represent negative measurements, which might be due to the non-linear galaxy bias, satellite fraction, or shot noise.}
    \label{fig: wgG ELG}
\end{figure}

We show the measured correlation functions for the DESI BGS g-g lensing in Fig.\,\ref{fig: wgG BGS}. The correlations are measured in 2 logarithmic bins in $0.5<\theta<80$ arcmin, with the statistical uncertainties calculated using jackknife re-sampling. We find that all three lensing surveys have strong g-g lensing signals, even for the current 1\% DESI data. The measurements are shown in blue dots (DECaLS), orange triangles (KiDS), and green squares (HSC), while the corresponding theoretical comparisons are shown in the blue solid curve, the orange dash-dotted curve, and the green dotted curve. From this figure, we find that the advantage of DECaLS is its large-scale cosmological information, with the highest S/N $\sim5.8$. This is due to DECaLS's significantly large overlap with DESI, reducing the cosmic variance. On the other hand, KiDS and HSC has larger S/N than DECaLS at small-scale, due to their higher source galaxy number density, which lowers the shape noise.

In this work we choose not to estimate the best-fit cosmology, as for DECaLS, there are some unaddressed potential systematics (as discussed in Sec\,\ref{sec DECaLS}), while for KiDS and HSC we do not want to harm the ongoing blinding efforts in the DESI collaboration (although for a larger catalog with the larger overlapped area). The theoretical estimations in Fig.\,\ref{fig: wgG BGS} and all the other similar figures in this work are based on the KiDS-1000 COSEBI $\Lambda$CDM cosmology with maximum posterior of the full multivariate distribution (MAP, \cite{Asgari2021}), which has $h=0.727$, $\Omega_{\rm b} h^2 =0.023$, $\Omega_{\rm c} h^2 =0.105$, $n_{\rm s} =0.949$ and $\sigma_8 =0.772$. We note the choice of other fiducial cosmology \citep{Planck2018I,Asgari2021,DESY3cosmo,HSC_Hamana2019} will give similar results for the current stage with DESI SV3. The linear galaxy biases are assumed following the descriptions of difference density tracers in Sec\,\ref{sec DESI}.

We note that the choice of 2 log-bins is limited by the 20 jackknife sub-regions \citep{Yao2020,Mandelbaum2006}, which is limited by: (1) the requirement of each jackknife sub-region is independent up to the largest scale we use (80 arcmin), and (2) the size of the overlapped region for KiDS and HSC ($\sim50$ deg$^2$). As the DESI survey expands, the available overlapped region will increase accordingly, resulting in increases in both the available number of sub-regions and the maximum angular scale we can measure. {Alternatively, we can use an analytical covariance (similar to Appendix \ref{sec apdx cov} but more tests need to be done) or simulation based covariance for future DESI data.} We also note in this work the inverses of the covariances are corrected \citep{Hartlap2007,Wang2020} due to the limited number of sub-regions.

As a demonstration of more angular binning, we use BGS$\times$DECaLS data to show the choice of 50 jackknife sub-regions and 5 angular bins, as in Fig.\,\ref{fig: wgG BGS DECaLS}. We show that with proper binning, more cosmological information can be extracted. The $\theta>\sim2$ arcmin measurements (the right 4 large-scale dots) agree with the linear bias assumption very well. In the future, with a larger overlapped footprint, more jackknife sub-regions can be used, so that more angular bins can be measured, either to increase the total S/N or to address any scale-dependent systematics.
We do see great potential for DECaLS from the above results, {although measurements will ultimately be limited by systematic errors.} 

We show the redshift distribution of the DESI LRGs and the three lensing surveys in Fig.\,\ref{fig: nz LRG}, requiring $z<0.6$ for the spec-$z$ LRGs and $0.7<z_p<1.5$ for the source galaxies. Similar to the BGS, more LRGs can be used when overlapping with DECaLS, while the available DECaLS source galaxies are less than in the other surveys. Since LRGs are generally distributed at higher $z$ than the BGS, we choose to increase the $z$-cut of the LRGs and the $z_p$-cut of the sources, resulting in reduced source galaxies compared with Fig.\,\ref{fig: nz BGS}. 
This figure shows the DECaLS source galaxies are more reduced (from 133k to 78k) as it is shallower than the other two.

The correlation measurements for the LRGs are presented in Fig.\,\ref{fig: wgG LRG}. At large-scale, the DECaLS signal is weaker than KiDS and HSC, but it still offers comparable S/N. At the small-scale, the S/N is dominated by deep surveys. The small-scale measurements are significantly higher than the theoretical predictions, due to LRGs being generally more massive than BGS, with stronger non-linear galaxy bias at such separations.

Furthermore, we study the g-g lensing measurements of the DESI ELGs. We show the redshift distribution of the DESI ELGs and the three lensing surveys in Fig.\,\ref{fig: nz ELG}, requiring $z<0.7$ for the spec-$z$ ELGs and $0.8<z_p<1.5$ for the source galaxies. The available number of galaxies is further reduced compared to BGS and LRGs, due to DESI ELGs being mainly distributed at $z>0.7$. And the high-z sources for DECaLS are significantly less than KiDS and HSC.

The correlation measurements of the ELGs are shown in Fig.\,\ref{fig: wgG ELG}. HSC appears to have the largest S/N at both large-scale and small-scale, and the S/N of DECaLS at large-scale is comparable to KiDS. 
{All three lensing surveys have small-scale measurements lower than the theoretical predictions, suggesting the low measurement is not a systematics of DECaLS. We suspect this might be due to shape noise, sample variance, or possibly non-linear galaxy bias. As when we go from large-scale to small-scale, the non-linear halo bias for less massive halos (for example the host halos for ELGs, see Fig.\,\ref{fig: wgG ELG}) tends to drop compared with its linear bias, while the non-linear halo bias tends to increase for the more massive halos (for example the host halos for the LRGs, see Fig.\,\ref{fig: wgG LRG}) according to Fig.\,1 of \cite{Fong2021}. The satellite galaxy fraction in the ELGs could also lead to a low amplitude at small-scale \citep{Niemiec2017,Favole2016,Gao2022}. These will require a higher S/N to test in the future. In this work, we only focus on large-scale ELGs measurement.}

\begin{table*}
	\centering
	\caption{We summarize the S/N of the DESI 1\% survey (SV3) g-g lensing results in Fig.\,\ref{fig: wgG BGS}, \ref{fig: wgG LRG} and \ref{fig: wgG ELG}, and forecast the {ideal} final S/N with full DESI, by rescaling the covariance based on the overlapped area, {and assuming DECaLS data can be well calibrated}. We note that the ELG measurements become negative sometimes, and therefore decide not to predict its final S/N. From this figure, we see that the advantage of DECaLS is at low-z (with BGS) and large-scale. We additionally present the possible bias in the forecasted S/N, namely $\Delta$S/N. It includes the contribution from the statistical error of the current measurement, and residual systematical bias from the data calibration. We use multiplicative bias $|m|\sim0.05$ \citep{Yao2020,Phriksee2020} and redshift bias $|\Delta z|\sim0.02$ \citep{Zhou2021} for DECaLS DR8,  $|m|\leq0.015$ and $|\Delta z|\leq0.013$ for KiDS \citep{Asgari2021}, and $|m|\leq0.03$ and $|\Delta z|\leq0.038$ for HSC \citep{HSC_Hikage2019}, to predict their systematical error in the forecasted S/N. We note the statistical contribution of $\Delta$S/N results from rescaling the $1\sigma$ error from Fig.\,\ref{fig: wgG BGS}, \ref{fig: wgG LRG} and \ref{fig: wgG ELG}, and is scale-independent and redshift-independent. The contribution from multiplicative bias $m$ is also scale-independent, while the contribution from redshift bias $\Delta z$ is weakly scale-dependent and redshift-dependent. In the table, we only show the $\Delta$S/N($\Delta z$) values corresponding to the BGS results at the large-scale.  }
	\label{tab forecast}
	\begin{tabular}{ c c c c c | c c c c | c c }
		\hline
		survey & SV3 overlap & \multicolumn{3}{|c|}{SV3 S/N [small-scale, large-scale]} & full overlap & \multicolumn{3}{|c|}{{ideal} forecast S/N [small-scale, large-scale]} &
		\multicolumn{2}{|c|}{forecast {potential bias} $\Delta$S/N} \\
		& [deg$^2$] & BGS & LRG & ELG & [deg$^2$] & BGS & LRG & ELG & statistical & systematical \\
		\hline
		DECaLS & 106 & [9.1, 5.8] & [3.5, 1.9] & [-0.3, 1.4] & $\sim9000$ & [83.8, 53.4] & [32.2, 17.5] & [N/A, 12.9] & $\pm9.2$ & $\pm5\%(m)\pm1.4\%(\Delta z)$ \\
		KiDS & 55 & [10.2, 3.9] & [8.7, 2.2] & [-1.1, 1.4] & 456 (DR4) & [29.3, 11.2] & [25.1, 6.3] & [N/A, 4.0] & $\pm2.9$ & $\pm1.5\%(m)\pm0.8\%(\Delta z)$\\
		HSC & 48 & [16.1, 4.3] & [10.6, 2.4] & [2.5, 2.6] & 733 (Y3) & [62.9, 16.8] & [41.4, 9.4] & [9.8, 10.2] & $\pm3.9$ & $\pm3\%(m)\pm1.6\%(\Delta z)$\\
		\hline
	\end{tabular}
\end{table*}

\subsection{Forecasts and Systematics}

We summarize our findings for the g-g lensing measurements from BGS (Fig.\,\ref{fig: wgG BGS}), LRGs (Fig.\,\ref{fig: wgG LRG}), and ELGs (Fig.\,\ref{fig: wgG ELG}) in Table\,\ref{tab forecast}. We see that DECaLS has its unique advantage in extracting cosmological information at large-scale and at lower redshift (when correlating with the DESI BGS). {Neglecting systematic errors for the moment, which will be dominant in practice, we give the forecast of the S/N with the complete DESI survey by re-scaling the covariance according to the overlapped area. This re-scaling assumes the covariance of the g-g lensing signal is dominated by the Gaussian covariance. Since we are extrapolating from small regions with significant boundary effects in our large-scale bin, this is only an approximation.}
We theoretically test the different components of the covariance in Appendix\,\ref{sec apdx cov} for your interest. The large-scale information of future DECaLS$\times$BGS can reach $>50\sigma$, which is stronger than most of the current g-g lensing data, 
and will be very promising in studying the current $S_8$ tension \citep{Hildebrandt2017,HSC_Hamana2019,HSC_Hikage2019,Asgari2021,Heymans2021,DESY3cosmo,DESY3model,DESY3data,Planck2018I}. 
The contribution from LRGs and ELGs, and possibly QSOs in the future, can also offer independent cosmological information.

\begin{figure}
	\includegraphics[width=\columnwidth]{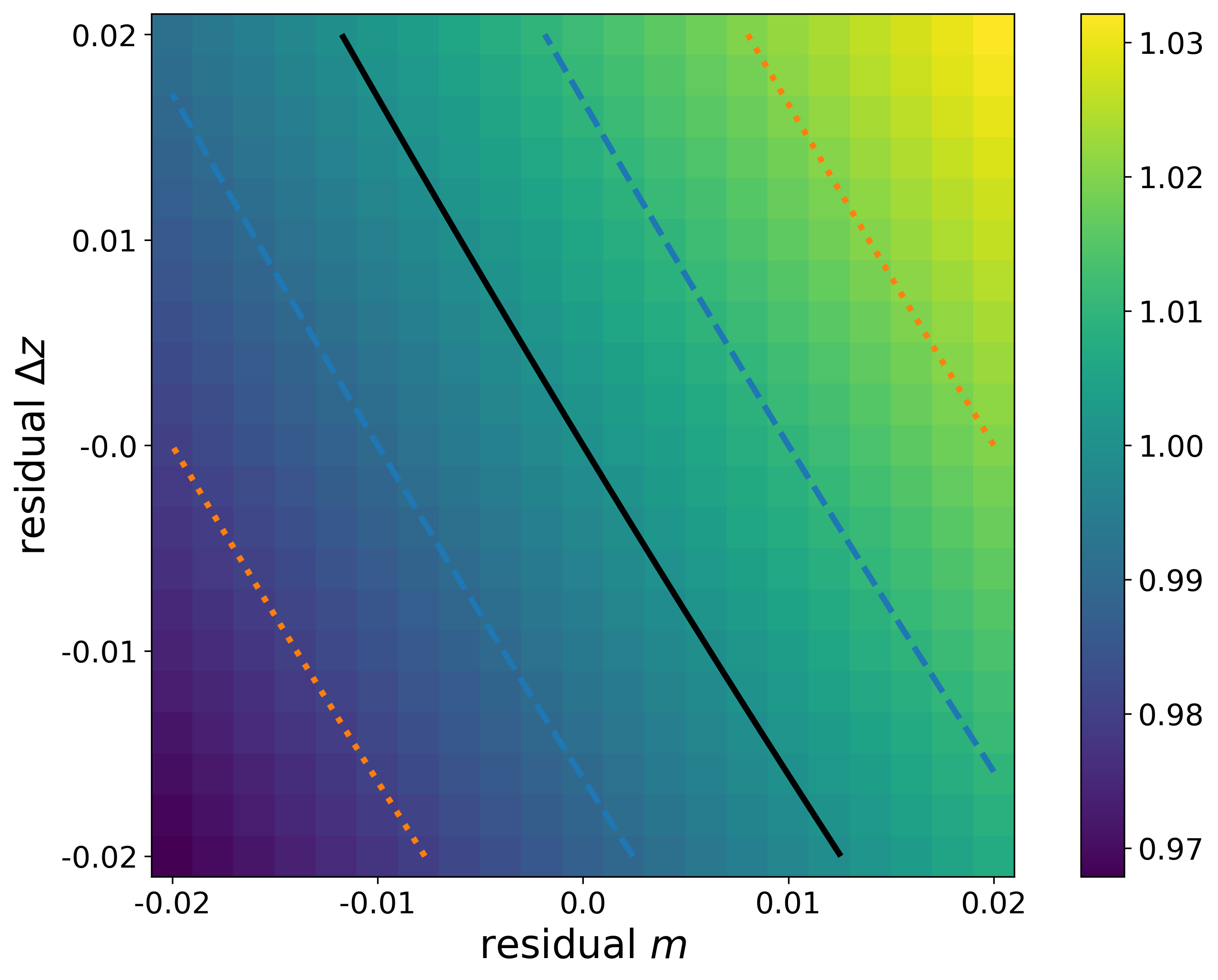}
    \caption{The impact of the residual shear multiplicative bias $m$ and the bias in the redshift distribution $\Delta z$. For different $m$ and $\Delta z$, we evaluate the resulting $w_{\rm bias}/w_{\rm true}$ at the large-scale of Fig.\,\ref{fig: wgG BGS}, \ref{fig: wgG LRG} and \ref{fig: wgG ELG} ($\theta\sim51$ arcmin) and show the ratio as the color map. The effect of $m$ is totally scale-independent, while the effect of $\Delta z$ is weakly scale-dependent, which can bring an additional $\sim20\%$ difference at maximum. We also show where the bias from $m$ and $\Delta z$ perfectly cancel each other (black solid curve), and the location where the net bias reaches $\pm0.01$ (blue dashed curve) and $\pm0.02$ (orange dotted curve).}
    \label{fig: sys bias}
\end{figure}

We note that the S/N predictions in Table\,\ref{tab forecast} ignored the potential bias from systematics, such as residual shear multiplicative bias $m$ and redshift distribution $n(z)$. The existence of the shear multiplicative bias $m$ will change the lensing efficiency from $q_{\rm s}$ to $(1+m)q_{\rm s}$ in Eq.\,\eqref{eq C^gG} and \eqref{eq q}. The bias in redshift distribution $\Delta z$ will change the redshift distribution for the source galaxies from $n_{\rm s}(\chi_{\rm s}(z_{\rm s}))$ to $n_{\rm s}(\chi_{\rm s}(z_{\rm s}-\Delta z))$ in Eq.\,\eqref{eq q}, {so that the whole redshift distribution is shifted towards higher-z direction by $\Delta z$}. For example, if we assume the residual multiplicative bias is $|m|\sim0.05$ (which is found for some DECaLS galaxy sub-samples as in \cite{Phriksee2020,Yao2020}), and enlarge the covariance to account for this potential bias, then the S/N of DECaLS$\times$BGS at large-scale will be reduced from $>50\sigma$ to $\sim20\sigma$. This is a huge loss of cosmological information, although $\sim20\sigma$ is still comparable to the $\sim11\sigma$ of KiDS-DR4 and $\sim 17\sigma$ of HSC-Y3. Therefore, we emphasize the importance of calibrating DECaLS data in a more precise way in the future for reliable cosmological measurements. We note the current measurements with DESI 1\% survey have S/N$\ll20\sigma$, therefore the impacts from such biases are still within the error budget. {The assumed systematics can enlarge the large(small)-scale uncertainties from $\sim17\%(\sim10\%)$ to $\sim18\%(\sim12\%)$.}

We further estimate the requirements on the DECaLS calibrations for precision cosmology. {We evaluate the fractional bias in the measured correlation function $w^{\rm gG}$, considering some residual multiplicative bias $m$ and redshift bias $\Delta z$, and present the results in Fig.\,\ref{fig: sys bias}.} To safely use the $\sim50\sigma$ data from the large-scale of DECaLS$\times$BGS, the residual multiplicative bias alone need to be controlled within $|m|<0.02$, and the mean of the redshift distribution of the source galaxies $\left<z\right>$ need to be controlled within $|\Delta z|<0.03$ on its own. {The net bias considering both $m$ and $\Delta z$ should be controlled in between the orange dotted curves in Fig.\,\ref{fig: sys bias}.} To safely use the cosmological information in both the large-scale and the small-scale, with overall S/N$\sim100\sigma$, we require the calibrations to have $|m|<0.01$ and $|\Delta z|<0.015$ individually, {while the net bias considering both $m$ and $\Delta z$ should be controlled in between the blue dashed curves in Fig.\,\ref{fig: sys bias}.}

We note that using tomography and combining g-g lensing measurements from different density tracers (BGS, LRGs, ELGs, and possibly QSOs in the future) can bring stronger S/N, so the requirements on the calibration terms will be more strict. However, these studies will require a much larger covariance, thus more jackknife sub-regions and much larger overlapped regions, which are beyond the ability of the current data size. We leave this study to future works.

\subsection{Shear-ratio}

\begin{figure}
	\includegraphics[width=\columnwidth]{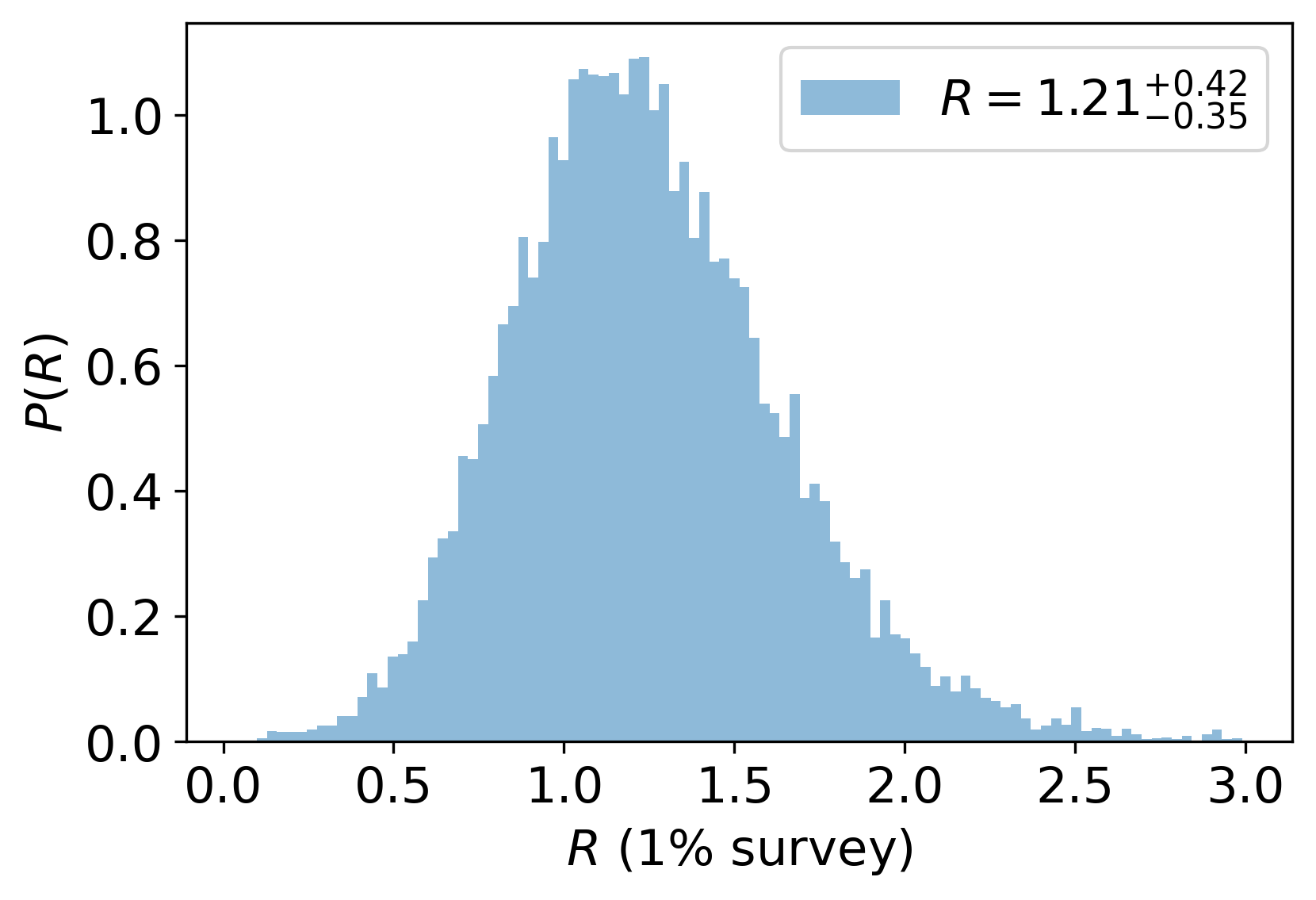}
    \caption{The {MCMC posterior PDF of the} shear-ratio measurements for BGS$\times$DECaLS using Eq.\,\eqref{eq constructed vector} and \eqref{eq constructed cov}. The galaxies are distributed as in Fig.\,\ref{fig: nz BGS}, with source galaxies split into $0.6<z_p<0.9$ and $0.9<z_p<1.5$. The constraint on the shear-ratio uses the two small-scale angular bins ($\theta<\sim5$ arcmin) as in Fig.\,\ref{fig: wgG BGS DECaLS}. The resulting $R=1.21^{+0.42}_{-0.35}$ agrees with the theoretical prediction between $1.13$ and $1.18$. When re-scaling the covariance to the final overlap of DESI$\times$DECaLS, the shear-ratio can be constrained as good as $\sigma_R\sim0.04$ when using the small-scale information, and $\sigma_R\sim0.03$ when using the full-scale.}
    \label{fig: R}
\end{figure}

Shear-ratio is a powerful tool to probe cosmology or test systematics \citep{Sanchez2021,Giblin2021}, and it is {insensitive} to many small-scale physics. As shown in Table\,\ref{tab forecast}, DECaLS$\times$DESI, especially for the BGS and LRGs, can offer very high S/N measurements at the small-scale. We take the BGS from the DESI 1\% survey as an example to study this topic.

The galaxy samples are distributed similarly to the BGS$\times$DECaLS $n(z)$ as in Fig.\,\ref{fig: nz BGS}, but in addition, the source galaxies are further split into two groups: $0.6<z_p<0.9$, and $0.9<z_p<1.5$. We calculated the corresponding correlations $w^{\rm gG}_1$ and $w^{\rm gG}_2$, and their ratio with $R=w^{\rm gG}_2 / w^{\rm gG}_1$, {following Eq.\,\eqref{eq constructed vector}, \eqref{eq constructed cov} and the description in Sec.\,\ref{sec shear-ratio}.}

The shear-ratio results are shown in Fig.\,\ref{fig: R}. Following the same angular binning as in Fig.\,\ref{fig: wgG BGS DECaLS} for the correlation calculations, we use the two small-scale angular bins with $\theta<\sim5$ arcmin, since the three large-scale bins are expected in the {direct 2-point cosmology} study, as described in Sec.\,\ref{sec wgG}. The current small-scale information can constrain shear-ratio at $R=1.21^{+0.42}_{-0.35}$, which is consistent with our theoretical prediction (using $R=w^{\rm gG}_2 / w^{\rm gG}_1$, Eq.\,\eqref{eq C^gG} and \eqref{eq w Hankel}) between $1.13$ and $1.18$. This small angular variation is due to the angular dependence in $P(k=\frac{\ell+1/2}{\chi},z)$ in Eq.\,\eqref{eq C^gG}, which is not fully canceled when taking the ratio using correlation functions. We note this weak angular dependence is small and can be easily taken into account in the theoretical predictions.

To predict the constraining power when full DESI finishes, we rescaled the covariance based on the overlapped area as in Table\,\ref{tab forecast}, and find the shear-ratio can be constrained at $\sigma_R=0.04$ with the small-scale information, which is not used in getting the $S_8$ constraint. Considering full information for the shear-ratio study, we can obtain $\sigma_R=0.03$. These statistical errors are comparable with the shear-ratio studies in \citep{Sanchez2021} with DES-Y3 data, {showing a promising future in using shear-ratio to improve cosmological constraint 
and/or to further constrain the systematics \citep{Giblin2021}. }

\subsection{Cosmic magnification}
\label{sec mag}

\begin{figure}
	\includegraphics[width=\columnwidth]{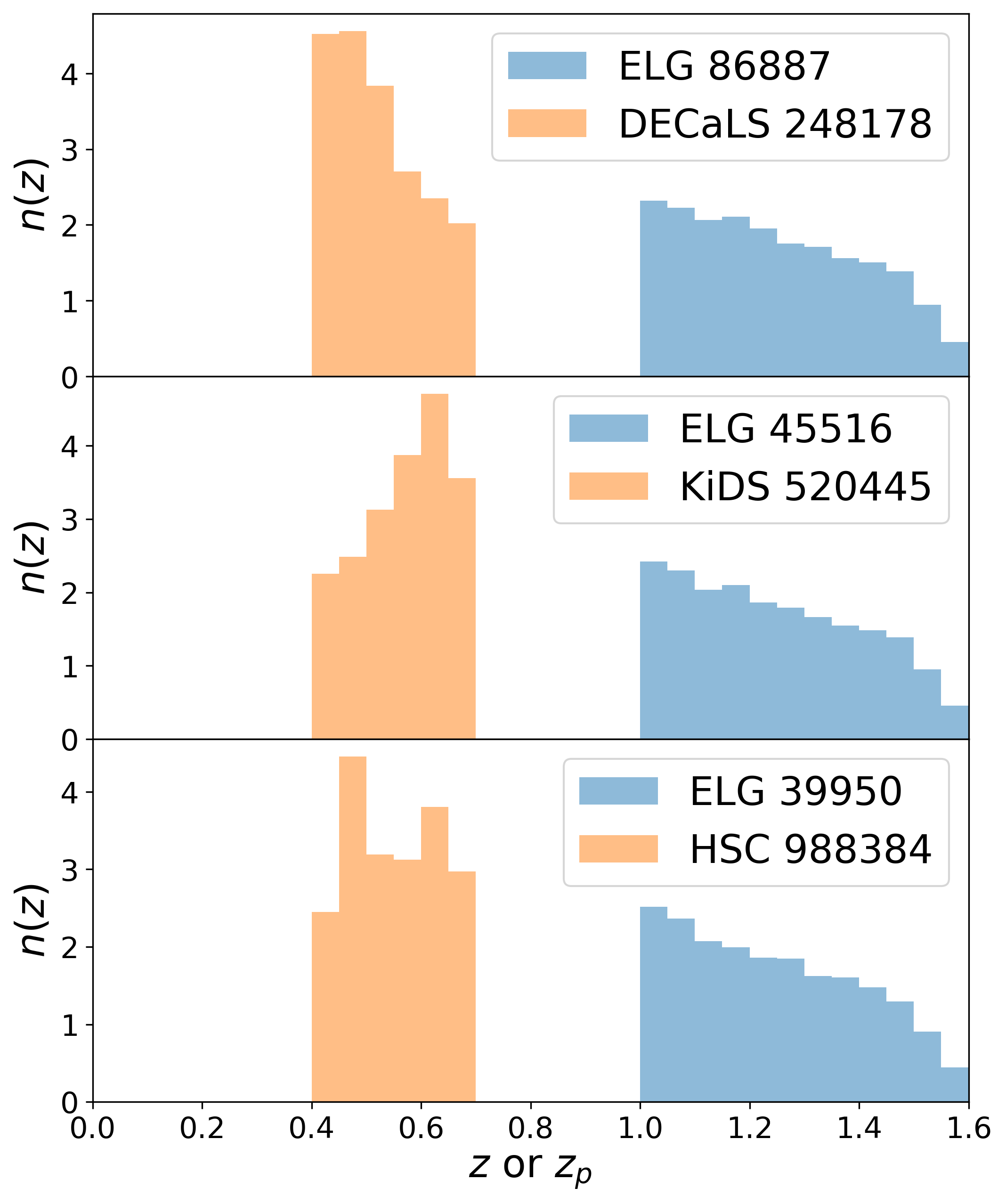}
    \caption{The redshift distribution for high-z ELGs ($1<z<1.6$) and low-z source galaxies ($0.4<z_p<0.7$) for magnification study. The choice of such a large redshift gap is to prevent potential leakage due to photo-$z$ inaccuracy. The numbers in the labels are the number of galaxies in the overlapped region.}
    \label{fig: nz ELG mag}
\end{figure}

We discussed that the ELG$\times$DECaLS results have low S/N in Fig.\,\ref{fig: nz ELG}, \ref{fig: wgG ELG} and Table\,\ref{tab forecast}, as the ELGs are mainly distributed at large-$z$, while the advantage of DECaLS is at low-$z$. On the other hand, this opens a window to the study of cosmic magnification by putting the ELGs at high-z and using shear from low-z DECaLS galaxies. We follow the methodology in \cite{Liu2021} and use galaxy samples distributed as in Fig.\,\ref{fig: nz ELG mag}. The DECaLS galaxies are located at a much lower photo-$z$ compared with the ELGs, as in the targeted shear-magnification correlation, the shear-density correlation exists as a source of systematics when even a small fraction of shear galaxies appear at higher-$z$ than the ELGs.

\begin{figure}
	\includegraphics[width=\columnwidth]{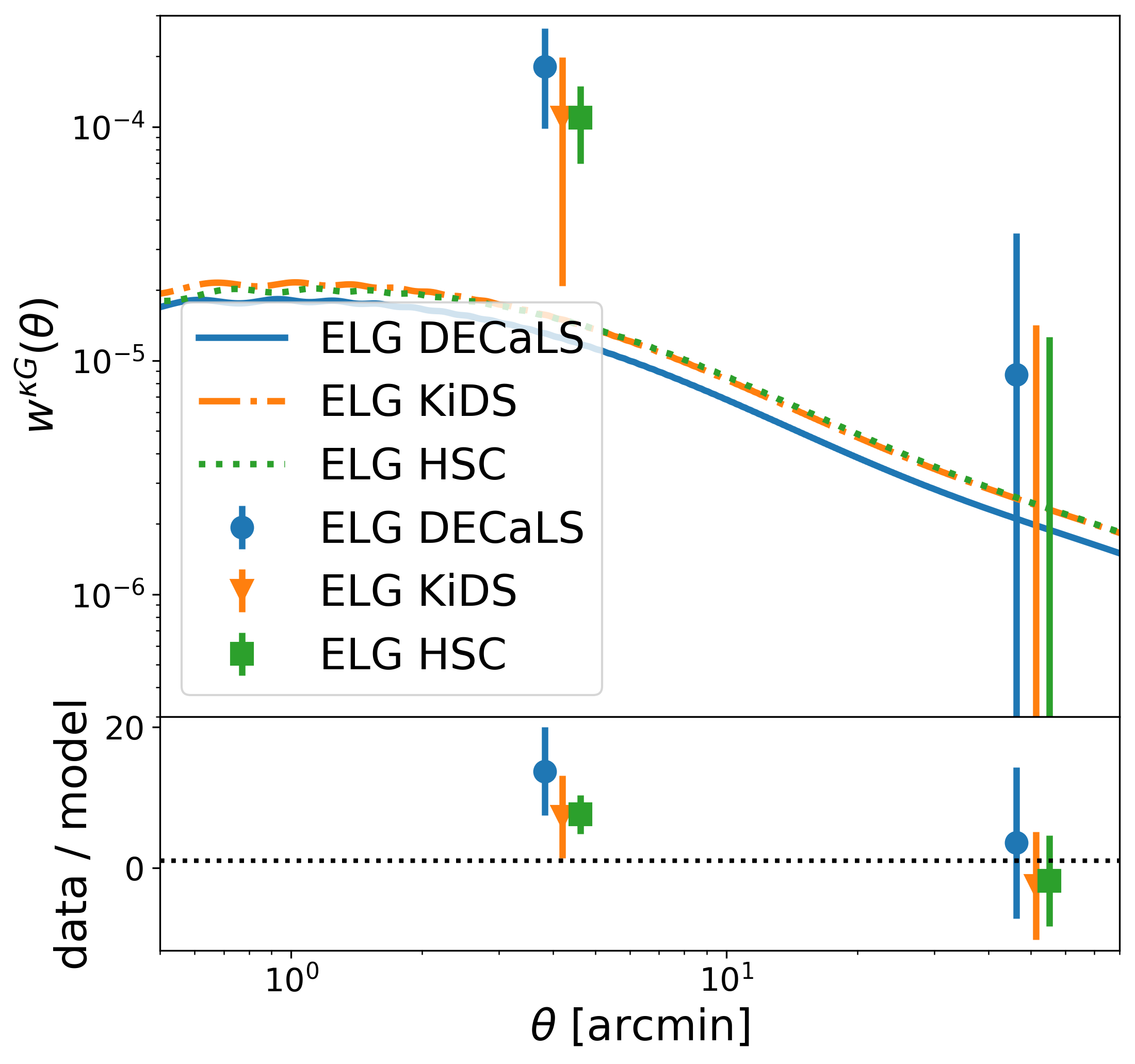}
    \caption{The magnification(ELGs)-shear correlation measurements, corresponding to the galaxy samples in Fig.\,\ref{fig: nz ELG mag}. The theoretical curves are based on Eq.\,\eqref{eq magnification}, assuming $g_{\mu,{\rm eff}}=1$ as a reference. The \{small-scale, large-scale\} detection significance for ELG$\times$DECaLS are \{2.2, 0.3\}, for ELG$\times$KiDS are \{1.2, -0.3\}, and for ELG$\times$HSC are \{2.8, -0.3\}. The negative values at the large-scale represent negative measurements, which might be due to shot noise, sample variance, or impact from systematics with negative values, like intrinsic alignment if there exists some photo-z outlier.}
    \label{fig: wgG ELG mag}
\end{figure}

The measurements are shown in Fig.\,\ref{fig: wgG ELG mag}. We find positive signals at the small-scale, and {null detections} at the large-scale, for all DECaLS, KiDS, and HSC. We tested the 45-deg rotation of the shear, resulting in consistency with 0 on all scales for all the source samples. Considering the similar calculation with eBOSS ELGs \footnote{\url{https://www.sdss.org/surveys/eboss/}} and DECaLS sources as a reference, we found the measurements are consistent with 0 on all scales, see Appendix\,\ref{sec apdx eBOSS} for details. In the measurements of Fig.\,\ref{fig: wgG ELG mag}, the {null detections} at the large-scale could be due to cosmic variance or some negative systematics such as intrinsic alignment. The positive measurements at the small-scale could be due to the targeted magnification signals, the cosmic variance, or photo-$z$ errors. We note to separate these different signals, either a stronger signal with clear angular dependencies or additional observables are needed to break the degeneracy.

As a further step, we present an effective amplitude fitting of $g_{\mu,{\rm eff}}$ for the magnification signals, following Eq.\,\eqref{eq C^muG}, in Table\,\ref{tab mag}. We find $\sim1\sigma$ {measurement} for KiDS and $\sim2\sigma$ {measurement} for DECaLS and HSC. Considering the ELG samples are quite similar as shown in Fig.\,\ref{fig: nz ELG mag}, and the three best-fit $g_{\mu,{\rm eff}}$-amplitudes are consistent, we evaluated the combined best-fit, achieving $\sim3\sigma$ significance. The covariance between different surveys is ignored for the combined estimation, as shot noise is more dominant in this case than the cosmic variance. Additionally, we find that by including shear galaxies from $0<z_p<0.4$, the significance of magnification detection drops, due to the low-z data having much weaker lensing efficiency as in Eq.\,\eqref{eq q}, and is mainly contributing noise. 

The fitting goodness of the reduced-$\chi^2$ (defined by the $\chi^2$ between the best-fit and the data, divided by the degree of freedom) is generally close to $\sim1$ for each case. This shows no significant deviation between the model and the data. The detected $\sim3\sigma$ positive signal can be either due to the cosmic magnification, or very similar stochastic photo-z outliers between the three lensing surveys. As DECaLS, KiDS and HSC have totally different photometric bands, photo-z algorithms, and training samples, we think the detected signals are less likely due to the similar photo-z outliers,
and more likely to be the cosmic magnification signal. Therefore, by assuming the combined best-fit of $g_{\mu,{\rm eff}}\sim6.1$ as the true value
and rescaling the covariance similar to Table\,\ref{tab forecast}, we expect $\sim10\sigma$ detection for DECaLS DR9,
which is very promising for a stage III lensing survey. By then, with a better understanding of the systematics such as IA and photo-$z$ outlier, these cross-correlations can bring very promising constraining power in studying cosmic magnification. {We can choose to: (1) cut a complete and flux-limited sample and compare it with the flux function; (2) try to use the given DESI completeness and flux function to find a relation of $g_{\mu,{\rm eff}}(\alpha)$ rather than $g_\mu=2(\alpha-1)$; (3) compare with realistic mocks to infer $g_{\mu,{\rm eff}}$; (4) add an artificial lensing signal $\kappa$ to real data and infer $g_{\mu,{\rm eff}}$ as a response $\partial\delta^{\rm L}_{\rm g}/\partial\kappa$, similar to MetaCalibration \citep{Sheldon2017,Huff2017}.}

\begin{table}
	\centering
	\caption{This table shows the best-fit amplitude $g_{\mu,{\rm eff}}$ for the cosmic magnification. The upper part corresponds to the results in Fig.\,\ref{fig: wgG ELG mag} for DECaLS, KiDS, HSC, and the combination of them (the ``all'' case). We find with the DESI 1\% survey, we can already detect cosmic magnification at $\sim3.1\sigma$ for the shear galaxies distributed at $0.4<z_p<0.7$, while the $z_p<0.4$ galaxies are mainly contributing noise as it corresponding lensing efficiency (Eq.\,\eqref{eq q}) is low. The degree of freedom is calculated as $dof=N_{\rm data}-N_{\rm para}$. We see no significant deviation between data and model as $\chi^2/dof\sim1$. 
	}
	\label{tab mag}
	\begin{tabular}{cccc} 
		\hline
		Case & $g_{\mu,{\rm eff}}$ & S/N & $\chi^2/dof$ \\
		\hline
		DECaLS $0.4<z_p<0.7$ & $10.6^{+5.2}_{-5.8}
$ & $1.8\sigma$ & 0.6/1 \\
        KiDS $0.4<z_p<0.7$ & $4.2^{+6.0}_{-5.7}
$ & $0.7\sigma$ & 1.3/1 \\
        HSC $0.4<z_p<0.7$ & $5.6^{+2.3}_{-2.3}$ & $2.4\sigma$ & 1.1/1 \\
        all $0.4<z_p<0.7$ & $6.1^{+1.9}_{-2.0}
$ & $3.1\sigma$ & 3.9/5 \\
        all $0<z_p<0.7$ & $5.3^{+2.0}_{-2.0}
$ & $2.7\sigma$ & 12.5/11 \\
        \hline
	\end{tabular}
\end{table}

\section{Conclusions}
\label{sec conclusions}

In this work, we study the cross-correlations between DESI 1\% survey galaxies and shear measured from DECaLS, one of the imaging surveys for DESI target selection. For the 1\% DESI data, DECaLS can have comparable performances compared with the main stage-III lensing surveys KiDS and HSC. More specifically, we measure the cross-correlations of DESI BGS/LRGs/ELGs $\times$ different shear catalog, shown in Fig.\,\ref{fig: wgG BGS}, \ref{fig: wgG LRG} and \ref{fig: wgG ELG}. We forecast the level of significance with full DESI data in Table\,\ref{tab forecast}. {Assuming systematic errors can be cleaned with high precision in the future,} we find the large-scale S/N could reach $>50 \sigma $ for DECaLS$\times$BGS, $>15\sigma$ for DECaLS$\times$LRG, and $>10\sigma$ for DECaLS$\times$ELG, which are very promising before the stage IV surveys come out.

We point out that the main difficulty in obtaining DECaLS cosmology is the calibrations for the systematics. In order to safely use the large-scale $\sim50\sigma$ information of BGS$\times$DECaLS, we need to achieve the minimum requirements on: (1) the multiplicative bias of $|m|<0.02$ and (2) the mean of redshift distribution $|\Delta z|<0.03$. To safely use the full-scale $\sim100\sigma$ data, we required $|m|<0.01$ and $|\Delta z|<0.015$ for future calibrations. The requirement could be even higher when combining different observables, but it will require a larger footprint than the 1\% survey for the study. These requirements are essential guides for future calibrations and studies on cosmology. 

To fully use the advantage of DECaLS, we further explored two promising observables, the shear-ratio, and the cosmic magnification. We show the current 1\% BGS data can constrain shear-ratio with $\sigma_R\sim0.4$, while the full DESI BGS can give $\sigma_R\sim0.04$ using only the small-scale information, as shown in Fig.\,\ref{fig: R}. Furthermore, weak detections of potential cosmic magnification are shown in Fig.\,\ref{fig: wgG ELG mag} and Table\,\ref{tab mag}. We discussed how the possible systematics can affect this signal in Sec.\,\ref{sec mag}. We also expect DECaLS to have a strong contribution ($\sim10\sigma$ detection) to future magnification studies, if the observed signals in this work are not due to fluctuations.

To summarize, DECaLS lensing is a very promising tool that can enrich the cosmological output of DESI. It will bring new cosmological information with its huge footprint. It has great advantages in the large-scale and the low-$z$ information, {after carefully addressing the systematics}. It will offer strong S/N for shear-ratio study, and good potential in measuring cosmic magnification. Careful calibrations of the shear and redshift distribution can result in very promising outcomes.

\section*{Acknowledgements}

We thank Xiangkun Liu, Weiwei Xu, and Jun Zhang for their helpful discussions. We thank Chris Blake, Daniel Gruen, and Benjamin Joachimi for their contribution during the DESI collaboration-wide review.

HYS acknowledges the support from NSFC of China under grant 11973070, the Shanghai Committee of Science and Technology grant No.19ZR1466600 and Key Research Program of Frontier Sciences, CAS, Grant No. ZDBS-LY-7013. PZ  acknowledges the support of NSFC No. 11621303, the National Key R\&D Program of China 2020YFC22016. JY acknowledges the support from NSFC Grant No.12203084, the China Postdoctoral Science Foundation Grant No. 2021T140451, and the Shanghai Post-doctoral Excellence Program Grant No. 2021419.
We acknowledge the support from the science research grants from the China Manned Space Project with NO. CMS-CSST-2021-A01, CMS-CSST-2021-A02 and NO. CMS-CSST-2021-B01.

We acknowledge the usage of the following packages pyccl\footnote{\url{https://github.com/LSSTDESC/CCL}, \citep{Chisari2019CCL}}, 
treecorr\footnote{\url{https://github.com/rmjarvis/TreeCorr}, \citep{Jarvis2004}}, 
healpy\footnote{\url{https://github.com/healpy/healpy}, \citep{Healpy_Gorski2005,Healpy_Zonca2019}}, 
matplotlib\footnote{\url{https://github.com/matplotlib/matplotlib}, \citep{Hunter2007}},
emcee\footnote{\url{https://github.com/dfm/emcee}, \citep{emcee}},
corner\footnote{\url{https://github.com/dfm/corner.py}, \citep{corner}},
astropy\footnote{\url{https://github.com/astropy/astropy}, \citep{astropy}},
pandas\footnote{\url{https://github.com/pandas-dev/pandas}},
scipy\footnote{\url{https://github.com/scipy/scipy}, \citep{scipy}},
dsigma\footnote{\url{https://github.com/johannesulf/dsigma}} 
for their accurate and fast performance and all their contributed authors.

This research is supported by the Director, Office of Science, Office of High Energy Physics of the U.S. Department of Energy under Contract No. DE–AC02–05CH11231, and by the National Energy Research Scientific Computing Center, a DOE Office of Science User Facility under the same contract; additional support for DESI is provided by the U.S. National Science Foundation, Division of Astronomical Sciences under Contract No. AST-0950945 to the NSF’s National Optical-Infrared Astronomy Research Laboratory; the Science and Technologies Facilities Council of the United Kingdom; the Gordon and Betty Moore Foundation; the Heising-Simons Foundation; the French Alternative Energies and Atomic Energy Commission (CEA); the National Council of Science and Technology of Mexico (CONACYT); the Ministry of Science and Innovation of Spain (MICINN), and by the DESI Member Institutions: \url{https://www.desi.lbl.gov/collaborating-institutions}.

The DESI Legacy Imaging Surveys consist of three individual and complementary projects: the Dark Energy Camera Legacy Survey (DECaLS), the Beijing-Arizona Sky Survey (BASS), and the Mayall z-band Legacy Survey (MzLS). DECaLS, BASS and MzLS together include data obtained, respectively, at the Blanco telescope, Cerro Tololo Inter-American Observatory, NSF’s NOIRLab; the Bok telescope, Steward Observatory, University of Arizona; and the Mayall telescope, Kitt Peak National Observatory, NOIRLab. NOIRLab is operated by the Association of Universities for Research in Astronomy (AURA) under a cooperative agreement with the National Science Foundation. Pipeline processing and analyses of the data were supported by NOIRLab and the Lawrence Berkeley National Laboratory. Legacy Surveys also uses data products from the Near-Earth Object Wide-field Infrared Survey Explorer (NEOWISE), a project of the Jet Propulsion Laboratory/California Institute of Technology, funded by the National Aeronautics and Space Administration. Legacy Surveys was supported by: the Director, Office of Science, Office of High Energy Physics of the U.S. Department of Energy; the National Energy Research Scientific Computing Center, a DOE Office of Science User Facility; the U.S. National Science Foundation, Division of Astronomical Sciences; the National Astronomical Observatories of China, the Chinese Academy of Sciences and the Chinese National Natural Science Foundation. LBNL is managed by the Regents of the University of California under contract to the U.S. Department of Energy. The complete acknowledgments can be found at \url{https://www.legacysurvey.org/}.

The authors are honored to be permitted to conduct scientific research on Iolkam Du’ag (Kitt Peak), a mountain with particular significance to the Tohono O’odham Nation.

\section*{Data Availability}

The data used to produce the figures in this work are available through \url{https://doi.org/10.5281/zenodo.7322710} following DESI Data Management Plan.

The inclusion of a Data Availability Statement is a requirement for articles published in MNRAS. Data Availability Statements provide a standardized format for readers to understand the availability of data underlying the research results described in the article. The statement may refer to original data generated in the course of the study or to third-party data analyzed in the article. The statement should describe and provide means of access, where possible, by linking to the data or providing the required accession numbers for the relevant databases or DOIs.



\bibliographystyle{mnras}
\bibliography{reference} 




\appendix

\section{Theoretical Covariance}
\label{sec apdx cov}

\begin{figure}
	\includegraphics[width=\columnwidth]{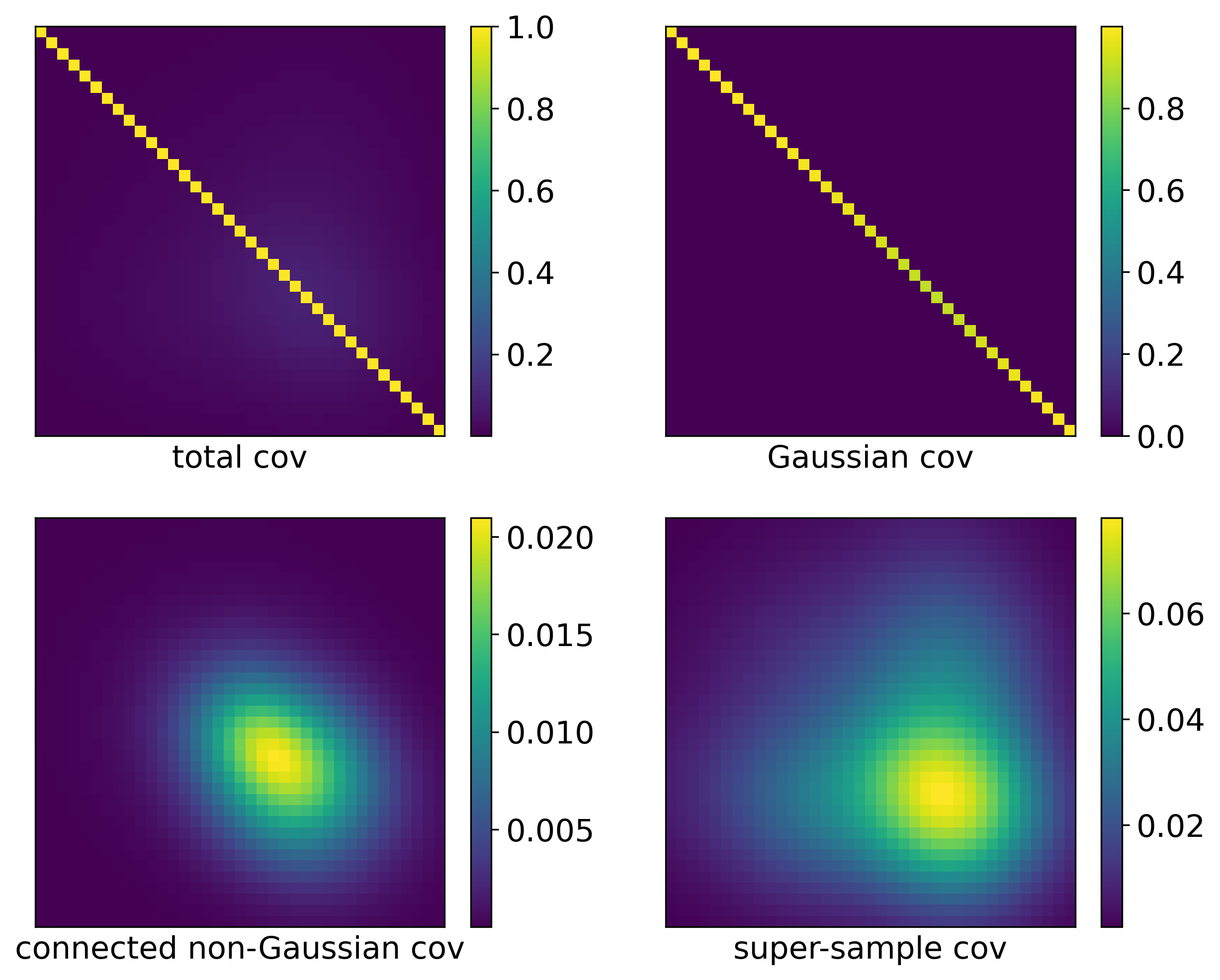}
    \caption{The theoretical covariance matrix (normalized, i.e. correlation coefficient) for the DECaLS$\times$BGS angular power spectrum, corresponding to the measurements in Fig.\,\ref{fig: wgG BGS DECaLS} and the DECaLS results in Fig.\,\ref{fig: wgG BGS}. It is clear the Gaussian component in the total covariance is much larger than the connected non-Gaussian component and the super-sample covariance component. }
    \label{fig: DECaLS X BGS cov}
\end{figure}

\begin{figure}
	\includegraphics[width=\columnwidth]{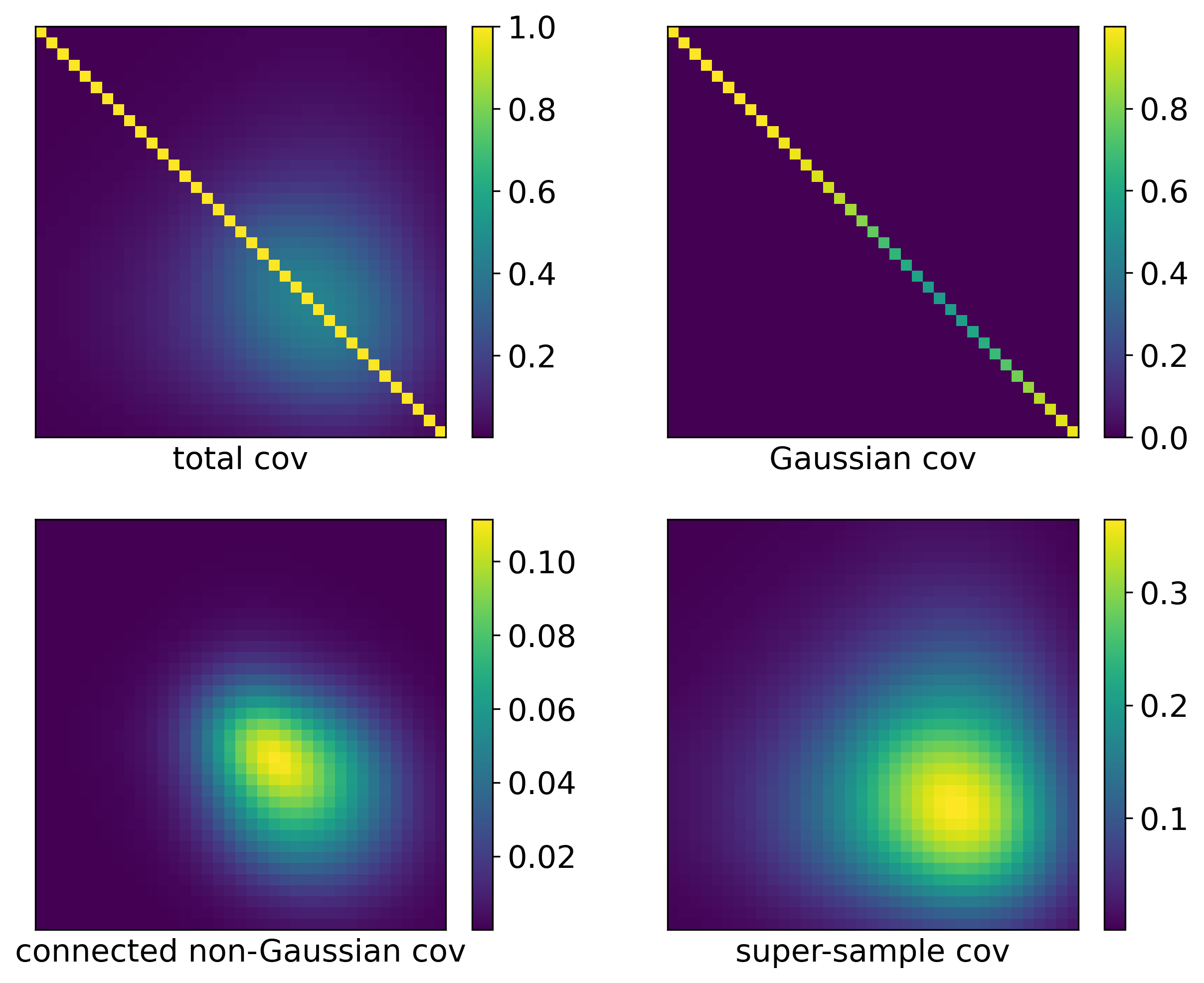}
    \caption{The theoretical covariance matrix (normalized, i.e. correlation coefficient) for the KiDS$\times$BGS angular power spectrum, corresponding to the measurements of the KiDS results in Fig.\,\ref{fig: wgG BGS}. The Gaussian component in the total covariance is still the dominant part. But the connected non-Gaussian component and the super-sample covariance component are relatively larger than Fig.\,\ref{fig: DECaLS X BGS cov} and are no longer negligible. }
    \label{fig: KiDS X BGS cov}
\end{figure}

We test the Gaussian covariance assumption being used in Table\,\ref{tab forecast} in this section. We use DECaLS$\times$BGS and KiDS$\times$BGS as examples, using the same galaxy number densities and redshift distributions as in Fig.\,\ref{fig: nz BGS}, and the same area as shown in Table\,\ref{tab forecast}. The angular power spectrum $C^{\rm gG}(\ell)$ is calculated within range $10<\ell<10000$, binned with $\Delta\ell=0.2\ell$, thus total $37$ angular bins. We follow the procedures in \cite{Joachimi2021} and divide the components into Gaussian covariance, connected non-Gaussian covariance, and super-sample covariance.

The Gaussian covariance is calculated by \begin{equation}
    {\rm Cov_G}(\ell_1,\ell_2)=\frac{\delta_{\ell_1,\ell_2}}{(2\ell+1)\Delta\ell f_{\rm sky}}\left[(C^{\rm gG})^2+(C^{\rm gg}+N^{\rm gg})(C^{\rm GG}+N^{\rm GG})\right],
\end{equation}
where $\delta_{\ell_1,\ell_2}$ is the Kronecker delta function; $C^{\rm gG}$, $C^{\rm gg}$ and $C^{\rm GG}$ are the galaxy-lensing, galaxy-galaxy, lensing-lensing angular power spectrum, respectively; $N^{\rm gg}=4\pi f_{\rm sky}/N_{\rm g}$ and $N^{\rm GG}=4\pi f_{\rm sky}\gamma_{\rm rms}^2/N_{\rm G}$ are the shot noise for $C^{\rm gg}$ and $C^{\rm GG}$, where $f_{\rm sky}$ is the fraction of sky of the overlapped area, $N_{\rm g}$ and $N_{\rm G}$ are the total number of the galaxies for the lens and source.

The connected non-Gaussian covariance \citep{Takada2004} is calculated by
\begin{equation}
    {\rm Cov_{cNG}}(\ell_1,\ell_2) = \int d\chi \frac{b_{\rm g}^2n_{\rm l}^2(\chi)q_{\rm s}^2(\chi)}{\chi^6}T_{\rm m}\left(\frac{\ell_1+1/2}{\chi},\frac{\ell_2+1/2}{\chi},a(\chi)\right),
\end{equation}
where $n_{\rm l}$ and $q_{\rm s}$ are the lens distribution and source lensing efficiency, $b_{\rm g}$ denotes the lens galaxy bias, $\chi$ denotes the comoving distance, same as those in Eq.\,\eqref{eq C^gG}; $T_{\rm m}$ is the matter trispectrum, calculated using a halo model formalism \citep{Joachimi2021}. We assume the NFW halo profile \citep{NFW1996} with a concentration-mass relation \citep{Duffy2008}, a halo mass function \citep{Tinker2008} and a halo bias \citep{Tinker2010}.

The super-sample covariance \citep{Takada2013} is calculated by
\begin{equation}
    {\rm Cov_{SSC}}(\ell_1,\ell_2) = \int d\chi \frac{b_{\rm g}^2n_{\rm l}^2(\chi)q_{\rm s}^2(\chi)}{\chi^6}
    \frac{\partial P_{\rm \delta}(\ell_1/\chi)}{\partial \delta_{\rm b}}
    \frac{\partial P_{\rm \delta}(\ell_2/\chi)}{\partial \delta_{\rm b}}
    \sigma^2_{\rm b}(\chi),
\end{equation}
where the derivative of $\partial P_{\rm \delta}/\partial \delta_{\rm b}$ gives the response of the matter power spectrum to a change of the background density contrast $\delta_{\rm b}$, while $\sigma^2_{\rm b}$ denote the variance of the background matter fluctuations in the given footprint. In this test, we use a circular disk that covers the same area as the given survey to calculate $\sigma^2_{\rm b}$.

The calculation is performed with the halo model tools in pyccl. We show the results of DECaLS$\times$BGS in Fig.\,\ref{fig: DECaLS X BGS cov} and KiDS$\times$BGS in Fig.\,\ref{fig: KiDS X BGS cov}. It is clear that the contribution from connected non-Gaussian covariance and super-sample covariance in DECaLS is negligible, so a Gaussian covariance can be fairly assumed for DECaLS in Table\,\ref{tab forecast}. The Gaussian covariance is still dominant in KiDS, however, the contribution from the other two is not negligible. Therefore, due to the small footprint, the forecasted S/N for KiDS and HSC in Table\,\ref{tab forecast} no longer scales exactly with the overlapped area.

We note that this test for different components of the covariance is only used to make an estimated comparison. Before using those covariances directly in the study, one needs to take care of the non-linear galaxy bias $b_{\rm g}$, the exact shape of the footprint that produces $\sigma^2_{\rm b}$, and build simulations to validate the accuracy of the theoretical covariance transferring from angular power spectrum to correlation functions as in \cite{Joachimi2021}. Therefore, we choose to stick with the data-driven jackknife covariance introduced in the main text, while we note that this effect could potentially reduce the forecasted S/N for KiDS and HSC in Table\,\ref{tab forecast}.

\section{eBOSS ELGs $\times$ DECaLS shear} \label{sec apdx eBOSS}

We show the cosmic magnification measurements using eBOSS ELGs $\times$ DECaLS shear, following a similar procedure as described in Sec.\,\ref{sec mag theory} and \ref{sec mag}. The overlapped area between eBOSS ELGs and DECaLS shear is $\sim930$ deg$^2$, which enables us to use 200 jackknife subregions and 5 angular bins, while we calculate the correlation in the angular range of $0.5<\theta<120$ arcmin, which is wider than Fig.\,\ref{fig: wgG BGS DECaLS}, see discussions in Sec\,\ref{sec wgG}.

\begin{figure}
	\includegraphics[width=\columnwidth]{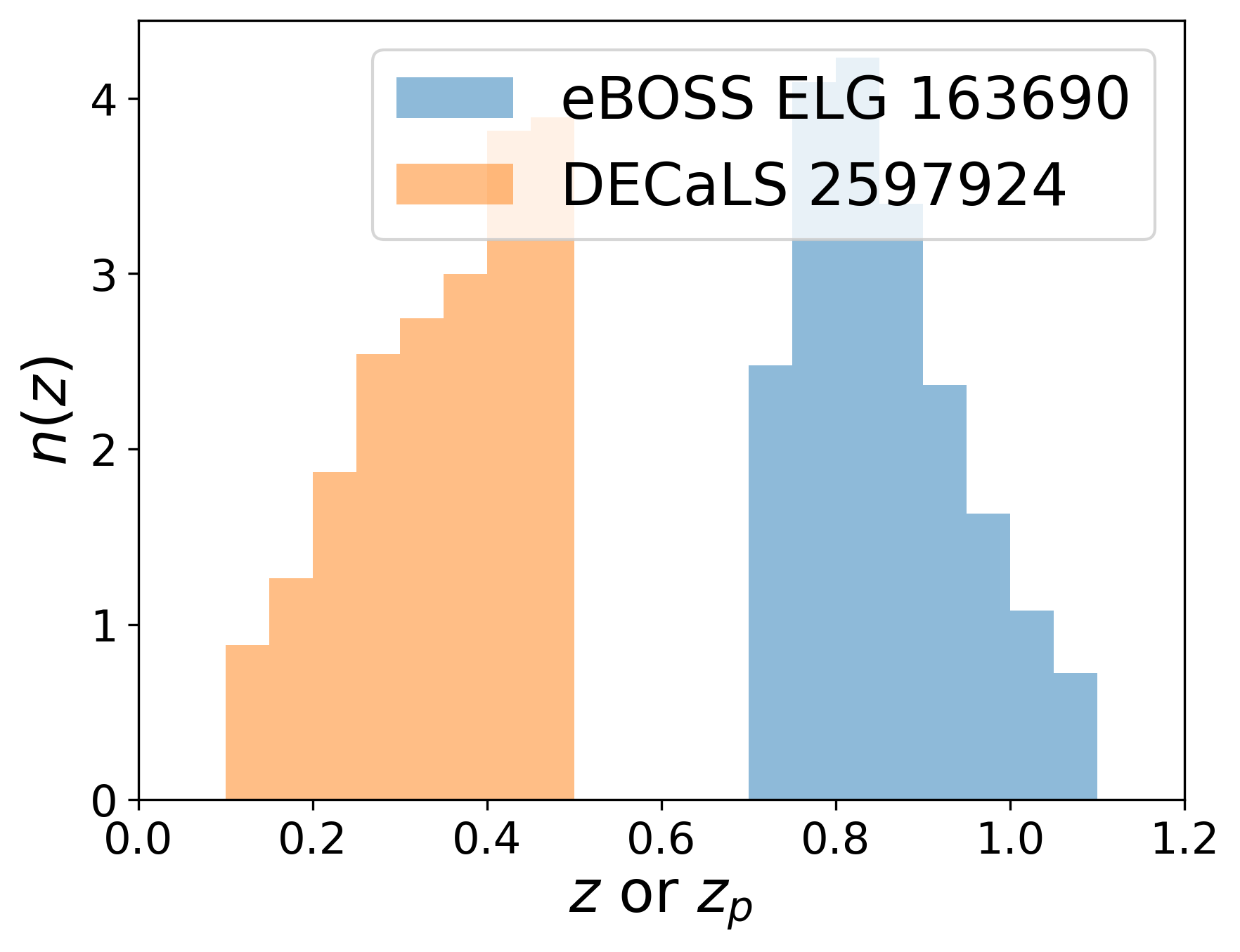}
    \caption{The galaxy redshift distribution for the eBOSS ELGs (blue) and photo-z distribution for DECaLS (orange). We use $0<z_p<0.5$ for DECaLS and $z>0.7$ for eBOSS ELGs. The redshift ranges are generally lower than Fig.\,\ref{fig: nz ELG mag} as eBOSS ELGs are at lower redshift than DESI ELGs.}
    \label{fig: eboss nz}
\end{figure}

\begin{figure}
	\includegraphics[width=\columnwidth]{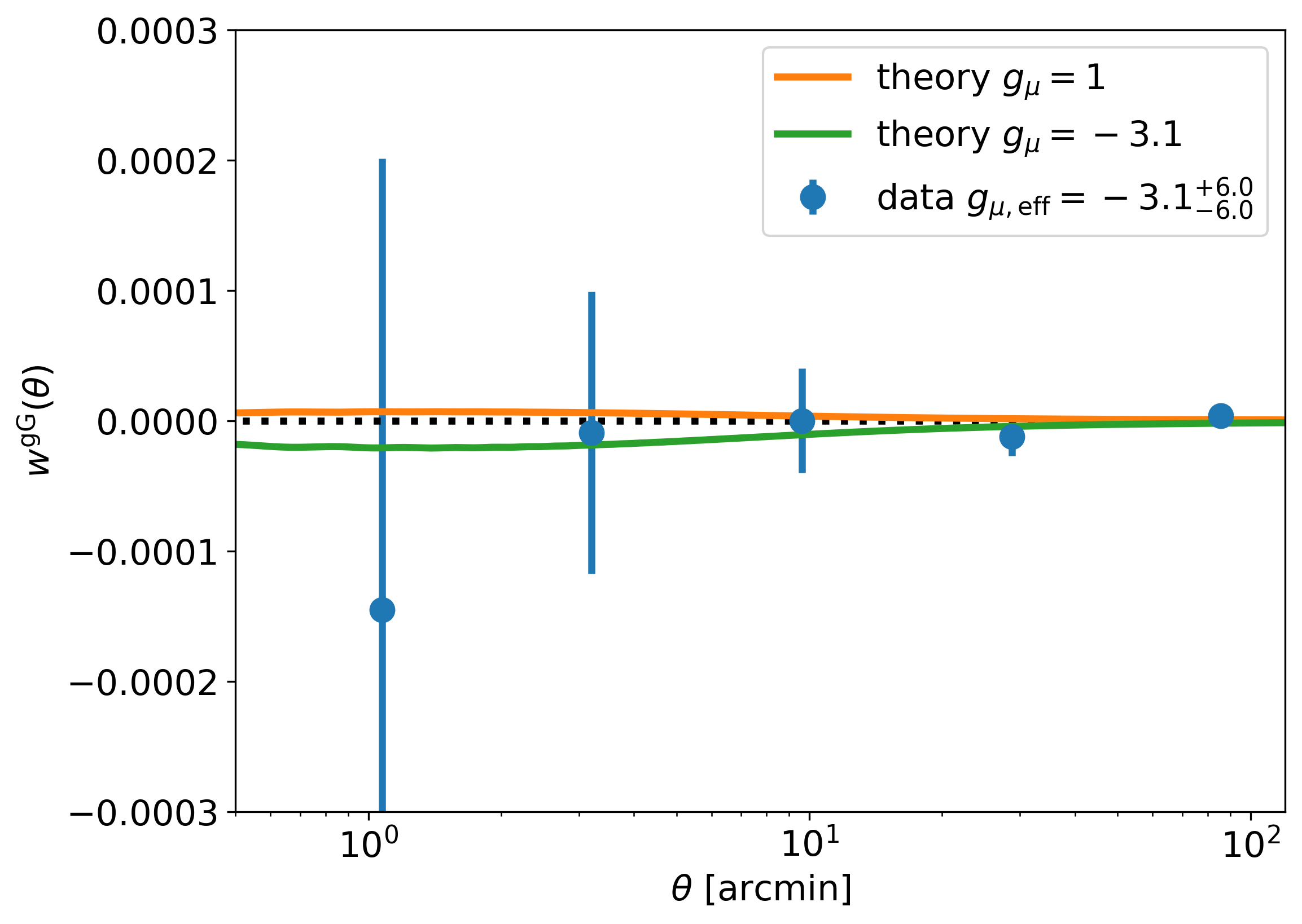}
    \caption{The magnification(ELGs)-shear correlation measurements for eBOSS$\times$DECaLS. Unlike Fig.\,\ref{fig: wgG ELG mag} for DESI, this measurement is consistent with 0.}
    \label{fig: eboss mag}
\end{figure}

In Fig.\,\ref{fig: eboss nz} we show the galaxy redshift distribution being used in this measurement. We see that the eBOSS ELGs are distributed at lower redshift compared with DESI ELGs in Fig.\,\ref{fig: nz ELG mag}, and more galaxies are used in this eBOSS measurement. The corresponding correlation function measurement is shown in Fig.\,\ref{fig: eboss mag}, which is consistent with 0. We think this is due to the fact that the galaxy number density for the eBOSS ELGs is much lower than the DESI ELGs, leading to a larger shot noise.





\bsp	
\label{lastpage}
\end{document}